\documentclass[]{aastex631}
\usepackage{amsmath}

\newcommand{\fermigbm}[1]{\textit{Fermi}-GBM {#1}}

\begin{document}

\title{Prompt GRB recognition through \textit{waterfalls} and deep learning}

\author[0000-0002-6548-5622]{Michela Negro}
\email{michelanegro@lsu.edu}
\affiliation{Department of Physics \& Astronomy, Louisiana State University, Baton Rouge, LA 70803, USA}

\author[0000-0003-3842-4493]{Nicol\'o Cibrario}
\affiliation{Istituto Nazionale di Fisica Nucleare, Sezione di Torino, Via Pietro Giuria 1, 10125 Torino, Italy}%
\affiliation{Dipartimento di Fisica, Università degli Studi di Torino, Via Pietro Giuria 1, 10125 Torino, Italy}

\author[0000-0002-2942-3379]{Eric Burns}
\affiliation{Department of Physics \& Astronomy, Louisiana State University, Baton Rouge, LA 70803, USA}

\author[0000-0001-9012-2463]{Joshua Wood}
\affiliation{ST12 Astrophysics Branch, NASA Marshall Space Flight Center, Huntsville, AL 35812, USA}

\author[0000-0002-0587-7042]{Adam Goldstein}
\affiliation{Science and Technology Institute, Universities Space Research Association, Huntsville, AL 35805, USA}

\author[0000-0001-5078-9044]{Tito \surname{Dal Canton}}
\affiliation{Université Paris-Saclay, CNRS/IN2P3, IJCLab, 91405 Orsay, France}

\begin{abstract}
Gamma-ray Bursts (GRBs) are one of the most energetic phenomena in the cosmos, whose study probes physics extremes beyond the reach of laboratories on Earth. Our quest to unravel the origin of these events and understand their underlying physics is far from complete. Central to this pursuit is the rapid classification of GRBs to guide follow-up observations and analysis across the electromagnetic spectrum and beyond.
Here, we introduce a compelling approach that can set milestone towards a new and robust GRB prompt classification method. Leveraging self-supervised deep learning, we pioneer a previously unexplored data product to approach this task: the GRB \textit{waterfalls}. 
\end{abstract}

\keywords{TBD}


\section{Introduction: A Holy Grail} \label{sec:intro}

Discovered serendipitously in the late nineteen-sixties \citep{klebesadel1973observations}, Gamma-Ray Bursts (GRBs) are among the most energetic events known in the universe \citep{burns2023grb}. Their discovery led to the beginning of gamma-ray astrophysics, which reaches the full potential when explored in concert with multiwavelength (across the electromagnetic spectrum) and multimessenger (adding information from gravitational waves, neutrinos, and cosmic rays) observations \citep{zhang2018physics}. The study of gamma-ray emission from GRBs, in particular, probes physics beyond extremes that can be achieved in terrestrial laboratories, enabling scientific conclusions in ultrarelativistic particle acceleration, the evolution of stars throughout the history of the universe, the production of heavy elements, fundamental physics, such as tests of gravity and the equation of state of neutron stars, and more \citep{burns2020neutron}. 

From an observational point of view, GRBs are promptly detected by gamma-ray burst monitors in space and look like a sudden rise in the photon count rate, followed by a typically slower decay and subsequent return to the background level between milliseconds and thousands of seconds in duration \citep{von2020fourth}. From an astrophysical perspective, GRBs are the manifestation of ultra-relativistic jets of plasma, highly collimated (typical jet half-opening angle is $\theta_{J}\leq5^\circ$) and directed towards us within a certain half-viewing angle $\theta_v$ which can be smaller (on-axis) or larger (off-axis) than $\theta_{J}$ \citep{salafia2015structure}. Physically, the prompt gamma-ray emission process is still an open question, as is the jet-launching mechanism, the structure of the magnetic fields, the nature of the particles involved (whether hadronic or leptonic), and the acceleration mechanism responsible for energizing them \citep{zhang2018physics}. Phenomenologically, GRBs exhibit diverse time and spectral characteristics and evolution, influenced by the progenitor event, viewing angle, and distance, necessitating tailored follow-up campaigns with telescopes spanning from radio to X-ray wavelengths, alongside focused searches for gravitational wave (GW) and neutrino signals. 

Known GRB progenitors are collapsars, a rare type of core-collapse supernovae \citep{galama1998unusual,cano2017observer}, typically associated with \textit{long} GRBs (duration $> 2$ seconds); binary neutron star (BNS) mergers, associated with \textit{short} GRBs (duration $< 2$ seconds \citep{fong2015decade}) unambiguously identified through the detection of gravitational waves predicted by General Relativity \citep{GW170817-GRB170817A}; extragalactic magnetar giant flares (MGFs) representing a small fraction of the observed \textit{short} GRBs and associated to local galaxies, with a characteristic luminosity several orders of magnitudes lower than other types of GRBs  \citep{hurley2005tremendous,Burns2021,trigg2023grb,mereghetti2024magnetar}. Much work remains to fully elucidate the origins of other types of observed or theorized GRBs. Notably, neutron star-black hole  mergers are likely to be a third short GRB progenitor \citep{mochkovitch1993gamma,foucart2014neutron}, but have not yet been observed in coincidence with candidates identified through GWs. Long GRBs can also be observationally separated into low-luminosity GRBs \citep{cano2017observer}, X-Ray Flashes \citep{barraud2005nature}, tidal disruption events \citep{burrows2011relativistic}, ultra-long GRBs \citep{levan2013new}, and long mergers \citep{rastinejad2022kilonova,levan2024heavy}. Each of which can correspond to different types of massive progenitors stars (e.g., long mergers are possibly from neutron star-white dwarf mergers \citep{zhong2023grb} and some ultra-longs may be from superluminous supernovae \citep{kann2024highly}) or result from a geometric effect where the collimated and relativistic jet is oriented away from Earth. 

The Fermi Gamma-ray Burst Monitor (\textit{Fermi}-GBM \citep{TheGBM}) on board NASA's Fermi Gamma-ray Space Telescope was designed to detect fast high-energy transient events. The instrument response is optimal in the energy range between 8 keV and 40 MeV. 
The GBM is composed of 14 scintillator crystals coupled to photo-multiplier tubes, 12 of which contain Sodium Iodide (NaI), covering the energy band 8--1000 keV, and the other two contain Bismuth Germanate (BGO) providing response between 0.2--40 MeV. The GBM data comes in different formats: data that are binned in time on-board and the unbinned event data. For the scope of this work we are interested in the unbinned Continuous Time-Tagged Event (CTTE) data, which are continuously downlinked (since late 2012) providing information of individual photons at 2 $\mu$s resolution and in 128 energy channels for each detector type. These data are the focus of offline sub-threshold analyses developed by the GBM team \citep{2015ApJS..217....8B, 2017ApJ...848L..14G, kocevski2018analysis}. 

\begin{figure*}
\includegraphics[width=\textwidth]{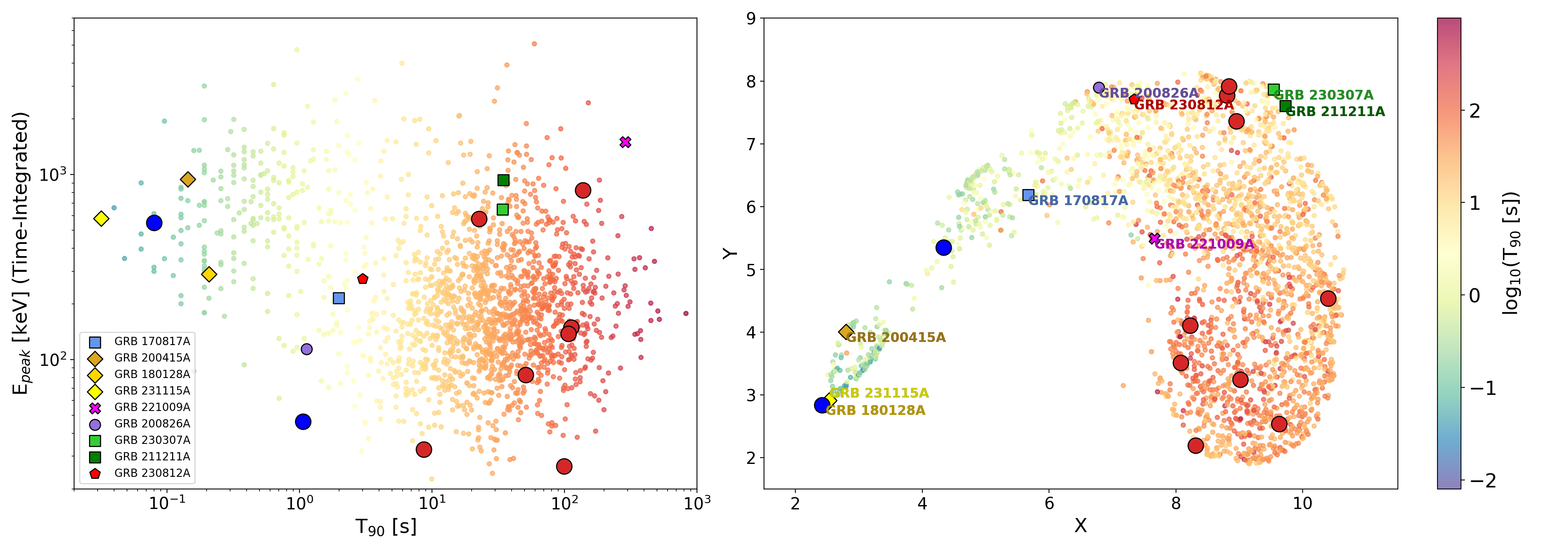}
\caption{\label{fig:BigFig} Left: the historical classification scheme (adapted from \citep{ahumada2021discovery}). Right: 2D GRB embedding. In both plots, the population of confirmed BNS mergers is reported as blue dots, while the population of confirmed collapsars are red dots. Additionally, we highlight the famous BNS counterpart of GW~170817, aka GRB\,170817A, shown as a blue square and the confirmed shortest collapsar GRB\,200826A, the purple circle. The long mergers GRBs\,230307A and 221112A are shown as green squares, while the yellow diamonds mark the MGF candidates.}
\end{figure*}

Linking prompt GRB detections to their origin not only yields critical insights into the physics governing the formation and propagation of ultra-relativistic jets, but it also facilitates targeted follow-up observations for comprehensive analysis. A basic and historical heuristic approach for separating GRBs into \textit{short} or \textit{long} classes is whether the duration is below or above a 2\,s threshold. A better method is fitting the bimodal distribution in 2D duration-spectral hardness space \citep{kouveliotou1993identification}. The classic 2D bimodal classification scheme utilized $T_{90}$ as a measure of duration and a hardness ratio, being the ratio of the flux between two predefined energy ranges.  A modern evolution on the prompt 2D bimodal classification is shown in Figure\,\ref{fig:BigFig} (panel (a)), which shows the $T_{90}$ and $E_{\rm peak}$, the energy at which the photon energy distribution peaks, which should better capture spectral hardness as it is avoids predefined energy ranges. 
This Figure also shows the limitation of such approaches. The confirmed BNS merger GRB\,170817A and the confirmed collapsar GRB\,200826A are both in the middle of the bimodal distributions. The MGFs are not obviously different than other short GRBs. Most interesting and confusing are the long mergers GRBs\,230307A and 221112A which solidly fall in the long class \citep{gompertz2023case,levan2024heavy}. Further, $E_{\rm Peak}$ 
is not available for $\sim$50\% of triggered bursts, as it is not always possible to constrain the curvature of the spectrum of a detected burst. 
The improper or ambiguous classification can arise in part because such approaches consider only a single measure of time and energy. The use of one or two integrated parameters also neglects the key spectral and temporal variations which likely provide additional information for the purposes of classification, as demonstrated by minimum variability studies (see, e.g., \citep{gompertz2023case, Camisasca2023, 2023ApJ...954L...5V}).

Here, we introduce a new tool that brings us a step closer towards the robust classification of GRBs, utilizing a self-supervised deep learning architecture and exploring a new set of data products that have not been previously investigated for this purpose: the GRB \textit{waterfalls} derived from the GRB prompt emission data captured by the Fermi-GBM.

\begin{figure*}
\includegraphics[width=\textwidth]{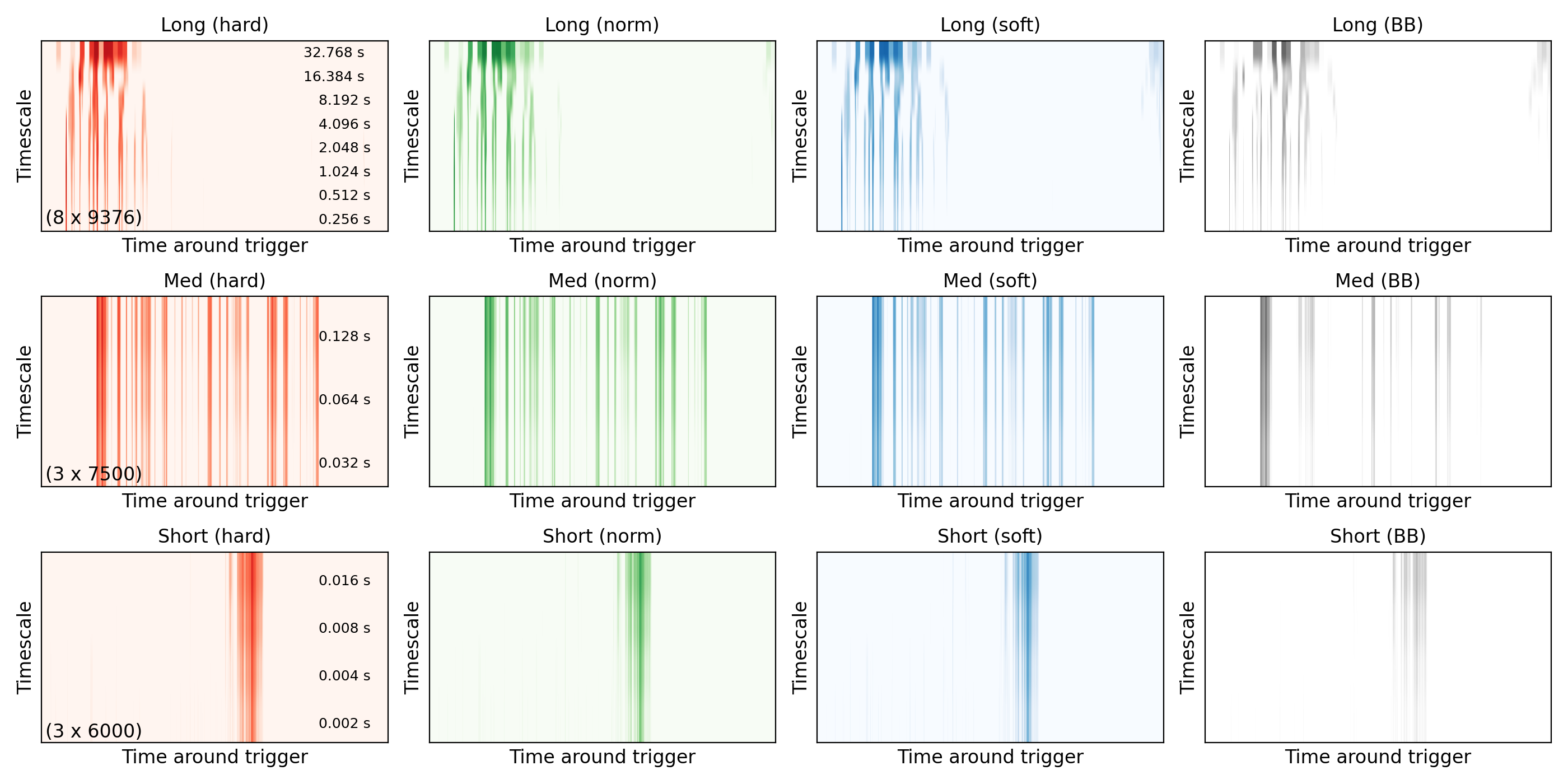}
\caption{\label{fig:input} Example of the set of inputs used to train the ConvAEs (MinVal=5). Top, middle, and bottom row show \textit{Long}, \textit{Medium}, and \textit{Short} timescales, respectively. The columns, from left to right, represent the different assumed spectral shapes \textit{Hard}, \textit{Norm}, \textit{Soft} and \textit{Blackbody}. For each timescale we report the size of the image in unit of pixels in the bottom corner of the `hard' column. More examples, in comparison with the reconstructed image by the ConvAE are provided in the SOM.}
\end{figure*}

The use of machine learning techniques for prompt GRB classification is not new. The key difference between previous studies based on prompt emission of GRBs \citep[see, e.g.,][]{Jespersen2020, Dimple2023, Chen2023, Garcia2023, Bhardwaj2023MNRAS, Dimple2024, Zhu2024, Nussle2024},  and our study is the input information passed to the self-trained deep learning algorithm. 
Therefore, we illustrate here the input features we feed to the neural network, while more details about the algorithm's architecture, a Convolutional Autoencoder (ConvAE), can be found in the Appendices.

The typical observable adopted as input for neural network algorithms in previous studies has been the GRB lightcurve, namely the rate of counts as a function of time around the GRB. Ref.~\citep{Jespersen2020} took this one step further by generating lightcurves for different energy bands within the range observed by \textit{Swift}/BAT. However, this type of input is limited by the choice of the temporal and energy binning, not fully capturing the spectral and time evolution of GRBs. This potentially causes important information loss. For example, the time binning used in Ref.~\citep{Jespersen2020} is not capturing variation on timescales finer that 64 ms \citep{2015ApJ...811...93G}, which appears to be an indicator of the GRB origin \citep{gompertz2023case, Camisasca2023, 2023ApJ...954L...5V}. Furthermore, that and other similar studies work in detector count space which is not directly tied to physical properties because of the non-linear and degenerate instrument response. A better approach is to work in deconvolved flux space, accounting for this response, to more directly utilize physical units to classify events. The use of multiple spectral templates should capture key behavior of the spectral evolution regardless of the true underlying spectrum. 
In an attempt to move toward classification based on physical properties, we developed a self-supervised neural network that takes advantage of 1) \fermigbm data; 2) the most advanced data processing developed for the GBM's GRB Targeted Search; and 3) a combination of convolutional neural networks and autoencoders to self-train the algorithm on $\sim$10\,years of observed events.

\section{Data and Analysis}
\textit{Data \& Analysis.}
We consider all GRBs that triggered the GBM since January 2013, once the downlink of CTTE began, until December 2023. This sample results in 2361 events used for the self-training (until May 2023) and 151 additional events used exclusively to evaluate and validate the trained model. Both training and evaluation sample includes GRBs with known progenitors, progenitor candidates, and events whose origin is unknown.

\textit{Waterfall plots}, or simply \textit{waterfalls}, have become more widely used in astronomy, generally showing how two-dimensional information changes over time. They can be generated for GRBs where the x-axis and y-axis give a time and duration, respectively, together specifying a source interval. In each interval, a color gradient represents a measure of brightness, or significance, in deconvolved flux space. In our case, it represents the log-likelihood ratio \citep{kocevski2018analysis,cai2021search}, which is determined by coherently fitting the normalization for an assumed spectral form to data from all GBM detectors. Multiple spectra can be fit separately, returning a different likelihood ratio at different time-scales depending on the phenomenology of the bursts. The Targeted Search relies on three template photon spectra, which are folded through each of the GBM detector responses. These templates, designated ``Soft'', ``Normal'', and ``Hard'',  consist of two Band functions \citep{band1993batse} and a power law with an exponential high-energy cutoff, respectively, and are intended to represent a range of GRB spectra observed by the GBM. For this work, we also include a ``blackbody'' spectrum which can capture some of the softest emission seen in time-resolved GRB studies \citep{GW170817-GRB170817A,burns2018fermi}. 

We scan the likelihood ratio around the trigger time, $T_0$, in time bins of decreasing size, starting from time scales of 32.768\,s, down to 2 ms. In particular we identify three sets of multiresolution timescales: \textit{Long}, \textit{Medium}, and \textit{short} timescales, organized in different resolution bins around the trigger time $T_0$ according to Table \ref{tab:timescales}. The information for each GRB is summarized as a 3 dimensional numpy array. The first dimension specifies which of the 12 entries, from the combination of 4 spectral templates and 3 timescales. The second dimension is the timescale and has size 8, and the third dimension the time. For cases where no data exists (i.e., indices above 3 for the timescale of the medium duration set) we enter values near 0. A set of these data products used as input for the algorithm is shown in Figure \ref{fig:input}. The GRB waterfalls encapsulate the core information relevant for GRB classification including duration, temporal variation, pulse structure, spectra, and how these parameters correlate and evolve. We tested different levels of background rejection, by setting a minimum value of likelihood ratio (MinVal = 0, 5, 10). For brevity, in the manuscript we report only the case of MinVal = 5, but the results for the other tests are similar and reported in the SOM, along with more details about the waterfall plots preparation. 

\begin{table}[b]
    \centering
    \begin{ruledtabular}
    \begin{tabular}{l|c|c|c}
         & \textbf{Timescales range} & \textbf{Num. of equal duration steps} & \textbf{Time around $T_0$} \\
         \hline
        \textit{Long} & (32.768 -- 0.256) s & 8 & ($-20$ -- $+280$) s\\
        \textit{Medium} & (0.128 -- 0.032) s & 3 & ($-10$ -- $+20$) s\\
        \textit{Short} & (0.016 -- 0.002) s & 4 & ($-2$ -- $+2$) s\\
    \end{tabular}
    \caption{Summary parameters of the GRB Waterfalls.}
    \label{tab:timescales}
    \end{ruledtabular}
\end{table}

The input described above feeds a self-supervised deep learning model that combines convolutional neural networks \citep{cnn} and autoencoders \citep{autoencoder}, known as a convolutional autoencoder \citep{convae}, that we denote ConvAE for brevity. Like typical autoencoders, ConvAE is composed of two main blocks: an \textit{encoder}, $e(.)$, which generates a compressed object $\hat{x}_e$ via a sequence of convolutional layers with different kernel sizes and strides; and a \textit{decoder}, $d(.)$, the reversed process. The weights of the model in each layer are optimized so that the output reproduces the input. The last layer of the encoder, which corresponds to the input of the decoder (the compressed or encoded state), is called \textit{latent space}. The general purpose of the ConvAE is to learn the best encoded representation of the events using an iterative optimization process. The learning process involves minimizing the loss function $L = ||x-d(e(x))||^2$, where \textit{x} denotes the input image, aiming to reduce discrepancies between the ConvAE output and the input image. For details on the ConvAE architecture, loss function and hyperparameters see Appendix~\ref{app:B}. 

We individually train three ConvAE, each tailored to a specific subset of time scales. Each input yields four distinct waterfall plots, corresponding to varying assumptions about energy spectra. After training, we combine the 10-dimensional latent space from each ConvAE in one 30-dimensional latent space (simply merging the linear arrays together, following their normalization). Each 30-D point in this space corresponds to a GRB. 

A dimensionality reduction is applied to the final latent space using the Uniform Manifold Approximation and Projection for Dimension Reduction UMAP, \citep{UMAP} algorithm to easily represent and interpret the results. The final outputs are shown in Figure \ref{fig:BigFig} (panel b), where we have set the UMAP to reduce the 30-dimension latent space into a 3D and 2D representations, allowing us to assess the results that we discuss in the next Section. We call these reduced data spaces, GRB \textit{embeddings}. In this manuscript we discuss the results based on the 2D embedding, as it offers an efficient static visualization and similar conclusions can be drawn by analyzing the 3D embedding. However, a visualization of the 3D embedding is offered as an interactive plot online (see appendix). 

Note that we optimize the ConvAE architecture and training parameters in a blind fashion, solely based on the comparison of the performance of the model. We make sure that the encoding-decoding process happens properly before revealing where the known GRBs fall by checking that the output correctly reproduces the input. We show a few examples of input-output comparison in Appendix~\ref{app:D}. Similarily, the UMAP parameters are optimized blindly based on the knonw and expected characteristics of GRBs (see Appendix~\ref{app:E})

\section{Results}
We now investigate the location of known GRBs in the embeddings. An interactive visualization of the GRB embeddings is available online\footnote{\url{https://nmik.github.io/SmartWaterfalls/plotly/grbs_umap_plotly.html}}. Here, we highlight the main findings, and more detailes can be found in the Appendices~\ref{app:B} and \ref{app:E}.

Looking at the 3D and 2D GRB embeddings, one can notice a bigger cluster, that we denote the \textit{head}, and a series of smaller clusters that together seem to define a \textit{tail}. 
Panel (b) of Figure \ref{fig:BigFig} illustrates some known GRBs in the 2D embedding. The 3D embedding is available for visualization at the online version of these plots.

Known collapsars with an observed associated supernova clearly populate the \textit{head} of the embeddings (red dots in Figure \ref{fig:BigFig}). We highlight GRB~230812A \citep{2023GCN.34391....1R}, a nearby collapsar that is part of our evaluation sample and it falls in the \textit{head} of the embedding despite having a quite short duration (T$_{90}\sim 3$ s).

In contrast, known BNS mergers fall sparsely along the \textit{tail} (blue dots in Figure \ref{fig:BigFig}). The short mergers are GRB~150101A \citep{burns2018fermi} and GRB~160821A \citep{2016GCN.19844....1P}, and GRB~170817A, corresponding to events for which the emergence of a SN has been observationally excluded at high confidence (that it, the bursts do not have a collapsar origin). The last BNS in the list is the famous GW-GRB event.

The confirmed BNS GRB~170817A (longer than 2 s) and the shortest confirmed collapsar GRB~200826A, both cross the standard 2\,s boundary in the standard classification method. The embeddings show these events are clearly separated, with the collapsar belonging to the `head' and the BNS to the `tail'. 
GRB~221009A, also known as the BOAT (The Brightest of all times)\citep{burns2023grb}, was a nearby collapsar and falls well within the head of the embeddings.

GRB~230307A \citep{levan2024heavy} and GRB~211211A \citep{rastinejad2022kilonova} are two \textit{long} GRBs ($\sim$ 100\,s) where the prompt emission is followed by a kilonova-like thermal transient, indicating a merger origin. Ideally these events would belong to a distinct cluster. Instead they fall in the \textit{head} cluster but are somewhat distinct from the shortest collapsar as they are marginalized to a rim close to the tail. Because the two long mergers were particularly bright, the current background fitting method used by the GBM targeted search will improperly track the burst, and the GRB waterfall plots may be affected by this, we investigated a possible dependence of the embedding distributions with the recorded fluence of each bursts. We find a negligible dependence of the embedding variables with the integrated flux, despite the fact that the latter spans several orders of magnitude. More details can be found in the Appendix.

So far, the GBM observed three GRBs with associated nearby host galaxies that are good MGF candidates: GRB~200415A \citep{roberts2021rapid}, GRB~180128A \citep{trigg2023grb}, GRB~231115A \citep{mereghetti2024magnetar}. GRB~200415A and GRB~180128A have similar duration, but the former has more power output in a shorter interval, which could explain why the algorithm locates them in different sub-clusters of the \textit{tail}. 
Note that our training sample does not include GRB~231115A, which is instead part of the evaluation sample. This GRB is almost co-spatial with GRB~180128A in the far tail of the 3D and 2D embeddings, which validates our model.

Overall, the algorithm seems to behave well in processing the validation sample, whose $T_{90}$ distribution within the final embedding follows that of the training sample.

Despite applying various unsupervised clustering algorithms, both hierarchical and density-based, we could not identify a method that reliably quantifies the probability of a GRB belonging to a particular group within the embedding. Many algorithms struggled with the physical expectation of highly variable cluster sizes and often maximized the number of clusters according to user specifications splitting the sample into similar sizes groups (see, e.g., Figure 1 in \citep{Dimple2024}), with some even producing different cluster counts when initialized with varying seeds. Furthermore, even if a reasonable clustering solution were found, fully characterizing the physical properties of these groups demands a thorough investigation of both the GRB prompt and afterglow emissions, drawing on dispersed and sometimes limited data available for each GRB. The work presented here establishes a foundation for what will be a multi-year study of these characteristics.

\section{Conclusions and Future prospects}
While the approach described in this manuscript demonstrates a promising improvement over the current state-of-the-art of standard and machine-learning GRB prompt classification studies, we have a path forward for additional improvements. These include changes to the construction of waterfall observables as well modifications to the machine learning method.

-- Implementing a polynomial model of the detector background rates computed from periods before and after the GRB as is done in spectral catalogs \citep{Poolakkil_2021}.  This will improve the accuracy of the waterfalls for bright and long duration GRBs, since the current background model used by the GBM targeted search averages over a 125 seconds window that includes the emission period, assuming that a putative signal is weak and short. 

-- We can account for the spacecraft motion when building the response matrix describing the source hypothesis in the targeted search. This will allow us to extend the input to emission timescales greater than $\sim$ minute. We expect this to lead to a greater differentiation of long GRB classes, especially the ultra-long GRBs. 

-- It might be interesting to add a spatial association variable that tests overlap with the Galactic plane. This could inform the neural network about features present in the time window around the trigger but unrelated to the GRB.  
This information can also be used to assign a probability of specific source intervals being a GRB (rather than a Galactic transient).

-- We can slightly modify the ConvAE to become a variational autoencoder \citep{vae}. This allows the latent space to be more regularized and structured, which may lead to an improved representation of the GRBs distribution. Furthermore, acknowledging that deep learning is a rapidly evolving field, we will further explore the state-of-the-art algorithms to potentially enhance the representation of the latent space.  

-- We plan to attempt semi-supervised clustering techniques, in which we inform the algorithm of the presence of a GRB of known progenitor class at a given location of the embedding.\\

We wish to acknowledge the GW-GBM working group for the fruitful discussions.

\bibliography{sample631}{}
\bibliographystyle{aasjournal}

\begin{figure}[b]
    \centering
    \includegraphics[width=1.\linewidth]{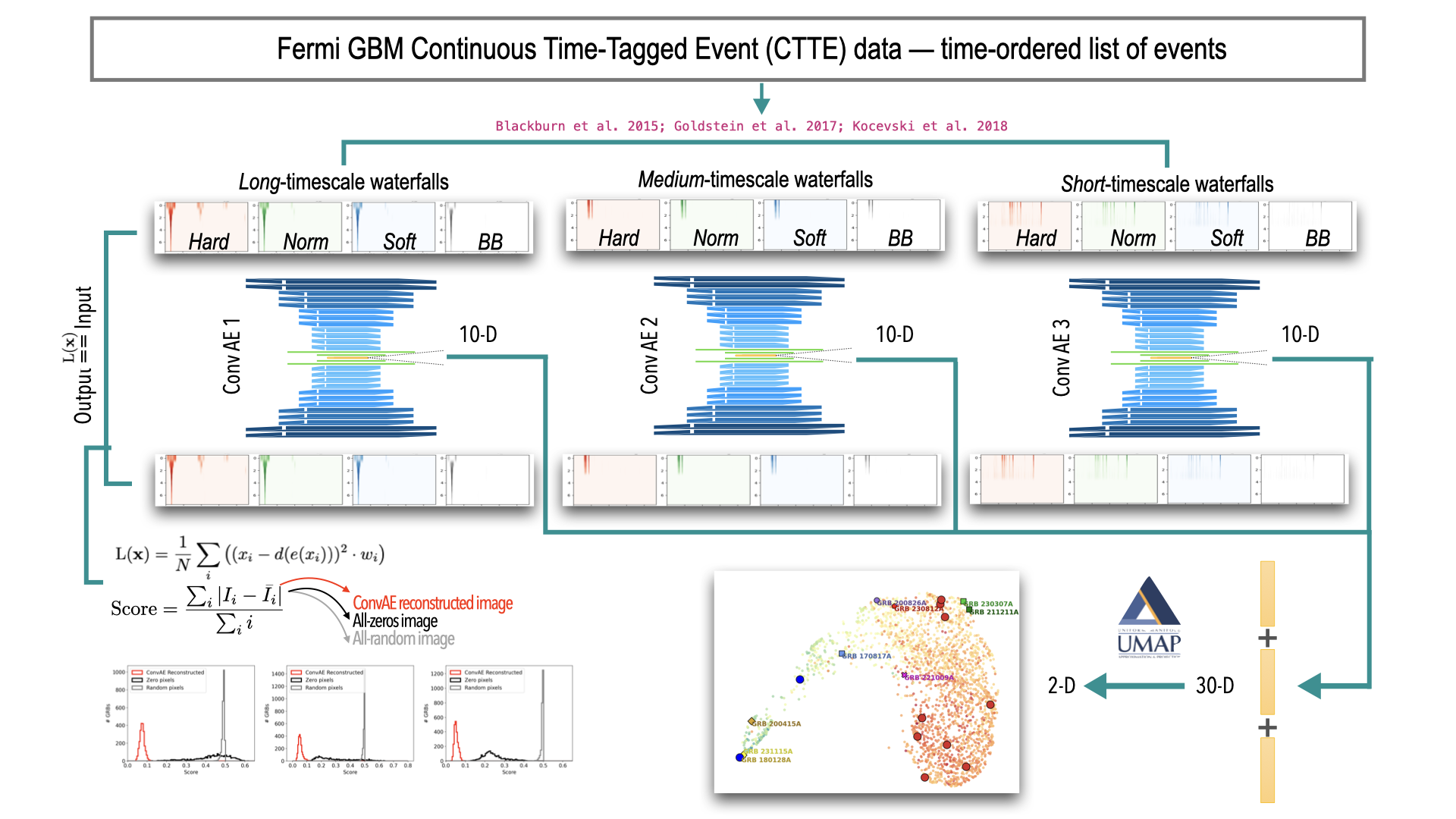}
    \caption{Flow chart of the procedure used to produce the 2D embedding (an analogous chart can be extrapolated for the 3D embedding). The three independent convolutional autoencoders are trained on waterfall plots of Fermi GBM CTTE data at different timescales (long, medium, short), where each GRB has four input images per timescale obtained assuming different spectra. Each ConvAE yields a 10-D latent space, and the three latent vectors are combined into a 30-D embedding before being projected via UMAP for clustering. Each autoencoder is optimized separately using its own reconstruction loss.}
    \label{fig:full_scheme}
\end{figure}

\appendix
In the following appendices, we provide additional details about the algorithm’s architecture. A comprehensive schematic of the entire design is illustrated in Figure~\ref{fig:full_scheme}.

\section{\label{app:A} GRB Waterfalls }

As mentioned in the main manuscript we make use of the \textit{waterfall plots} developed by the \fermigbm{} Team for the sub-threshold targeted transient search. The search looks for a signal above a noise threshold in the time sequence of the counts in all the GBM detectors, around the time of a trigger. The search is run over a predefined total temporal window of interest. The search is run over a range of timescales, $\Delta T$, in order to quantify signal strength over a variety of durations. Within each timescale the source interval steps forward by $\Delta T / n_{\rm steps}$, running the search through overlapping source intervals in order to ensure the phase of our binning does not miss significant emission. The $\Delta T=2$\,ms timescale utilizes $n_{\rm steps} = 4$ while all other intervals use $n_{\rm steps}=8$. For example, the $\Delta T=1.024$\,s timescale steps forward in $0.128$\,s steps.

We run the search over the full sky utilizing a rough spatial grid. We convolve a fixed spectral function with the detector response at each position and determine the signal strength via a log likelihood ratio calculation, described in \citep{2015ApJS..217....8B}. We marginalize over the log likelihood ratio values across the spatial grid in order to determine a single value for the significance of a transient during that source interval (assuming the spectral form).

\begin{figure*}
     \includegraphics[width=\textwidth]{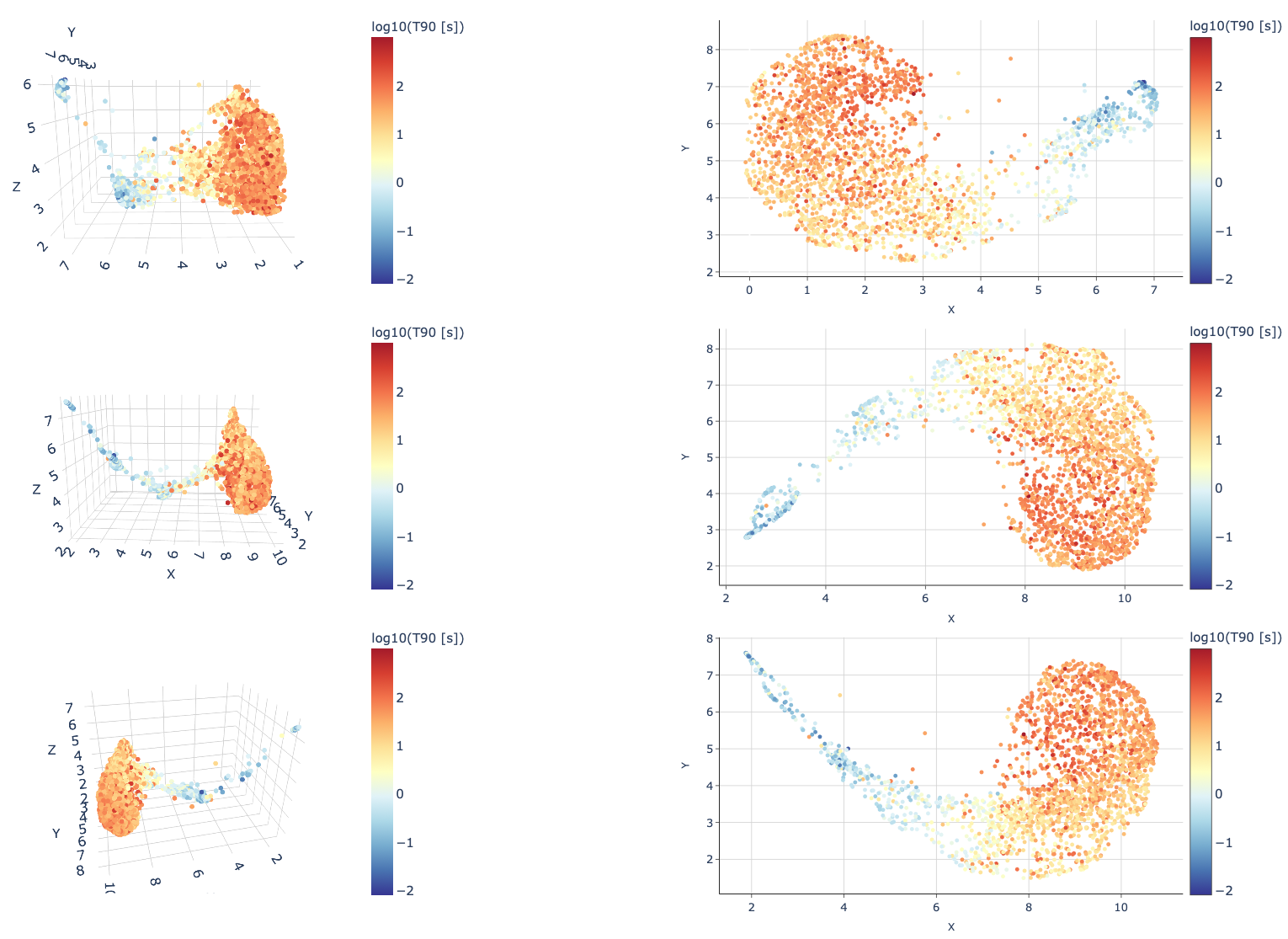}

     \caption{GRB 3D (left) and 2D (right) embedding for MinVal=0 (top), MinVal=5 (middle) and MinVal=10 (bottom).}
     \label{fig:minval}
\end{figure*}

We used different minimum values of Likelihood Ratio (MinVal) for background rejection: MinVal = 0, 5, 10. We note that the denominator of the log likelihood ratio is arbitrary (as this does not affect its use as a false alarm rate ranking statistic, the primary purpose of the search), so these MinVal thresholds are also arbitrary, with the selection of 10 being empirical. In the main manuscript we report the results for MinVal = 5, while here we report the remaining cases. Figure~\ref{fig:minval} shows the comparison of the 3D and 2D embedding shapes for different MinVals. The main structure of the latent space is similar, symptomatic of an healthy functioning of the algorithm. 

For each GRB the set of input waterfall plots is normalized to the maximum log-likelihood value among all the waterfall plots of that given GRB, so that each log-likelihood value, for each GRB is defined as:
\begin{equation}
    \mathcal{L}^{i'}_{\rm pix} = \frac{\mathcal{L}^i_{\rm pix}}{\mathcal{L}^{\rm all~waterfalls}_{\rm max}} 
\end{equation}
where $i$ is the $i^{th}$ pixel in any waterfall image of a given GRB. This allows us to have all images ranging between 0 and 1, which is typically advantageous for machine learning applications.

\section{\label{app:B} The Embeddings}
Interactive 3D and 2D plots for the different MinVal values are visible at the following links:
\begin{itemize}
    \item ${\rm MinVal}=0$:\\ \url{https://nmik.github.io/SmartWaterfalls/plotly/grbs_mv0umap_plotly_v0.html}
    \item ${\rm MinVal}=5$:\\ \url{https://nmik.github.io/SmartWaterfalls/plotly/grbs_mv5umap_plotly_v0.html}
    \item ${\rm MinVal}=10$:\\ \url{https://nmik.github.io/SmartWaterfalls/plotly/grbs_mv10umap_plotly_v0.html}
\end{itemize}
In each of the images linked above, we marked the main GRBs discussed in the last section of the main manuscript.
Additionally we report the known CCSNs and the known BNSs in red and blue circles respectively. The three rows of plots show how different variables are distributed in the embeddings with different color scales. The flux, $F_{64}$, (top row), the $T_{90}$ (middle row), and the energy peak (bottom row).

As expected, the single best determinant of burst separation is the $T_{90}$, with the bulk of bursts in the largest group matching the long class and the extended tail matching short bursts. However, there are exceptions to the basic rule, showing the algorithm is handling additional information. Similarly, owing to detection methods, the time-integrated fluence is lower for the group of short events than the larger long group. Within the large blob contains structure based on fluence. While $E_{peak}$ is not measured for all bursts, the tail contains generally higher values. Lastly, additional information is being captured beyond the parameters captured in the GBM catalogs, as shown by the isolation of the long mergers.


\section{\label{sec:som-ae} The ConvAE architecture}

\begin{center}
\begin{figure}[b]
    \includegraphics[width=\textwidth]{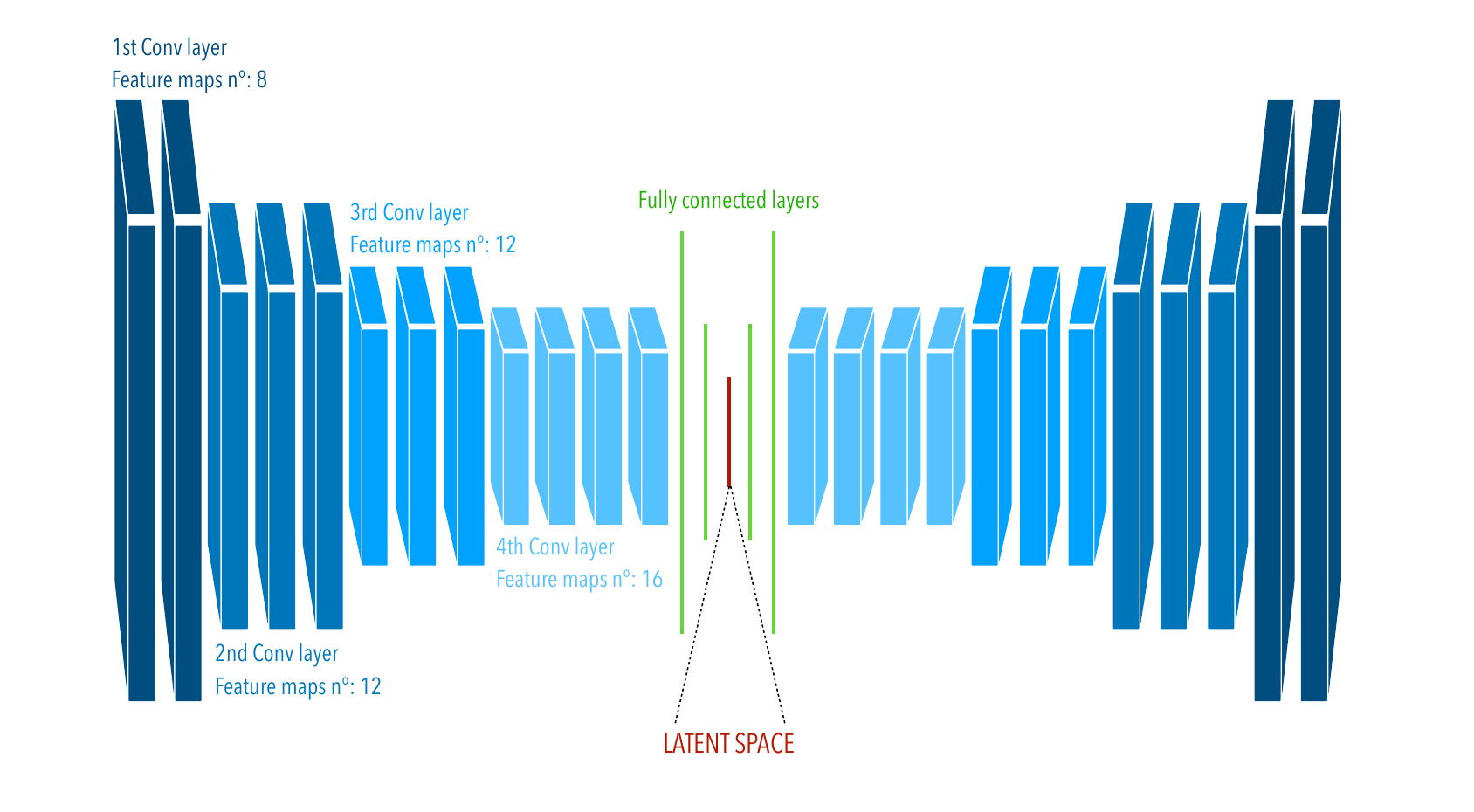}
    \caption{Simplified graph of the structure of our ConvAE. The input images are passed through a series of convolutional layers, each time increasing the number of features as marked in the figure. The dimensionality reduction at each step of the convolutional layers are achieved by changing the kernel size and the stride. The kernel sizes of the four convolutional layers varies in the three ConvAE for the different timescales. In order from the outer to the inner layer of the encoader, for Large timescales, we have kernels (4x4), (1x3), (2x3), (2x3), and strides (2x2), (1x2), (1x2), (1x2); for Medium timescales we set kernels (2x2), (1x3), (1x2), (2x3) and strides (1x2), (1x3), (1x2), (1x2); for Short timescales we set kernels (2x2), (1x2), (2x2), (2x2) and strides (1x2), (1x2), (1x2), (1x2). The inner structure of the algorithms sees 3 fully connected layers, that reduce the dimensional down to a 10-element linear array (the latent space). The final latent space is a 30-element array obtained by summing the 10-element latent spaces of the individual ConvAE trained on the different timescales.}
    \label{fig:convae}
\end{figure}
\end{center}

In this section we present the structure of our Convolutional Autoencoder (ConvAE). We refer to Section `Data Analysis' of the manuscript and Appendix~\ref{app:A} for the detailed description of the input. The encoder is a simple convolutional neural network, which does not present pooling layers. The architecture consists in four convolutional blocks, each generating respectively 8, 12, 12 and 16 feature maps, with kernel sizes and strides which differ accordingly to the images input size. The last feature maps are converted to a 1-dimensional array, which is passed through a fully connected network with two hidden layers (512 and 64 neurons respectively), with a final 10-dimensional latent space. From here, the decoder part begins: it mirrors the encoder following the same structure as just described from the bottom up. The convolution operator is replaced by a transposed convolution operator, which can be seen as a gradient of the convolution operator respect to its input \citep{deconvolution}. This is illustrated in Figure \ref{fig:convae}.

The training process optimizes the input-output matching by minimizing the loss function. We adopt a \textit{Mean Squared Error} (MSE) loss, but based on the MinVal value and the timescale of the images, the loss function is modified to weight pixels with specific intensities during model training. This adjustment is particularly helpful as many pixels’ intensity is zero, which can disproportionately affect the loss computation. By increasing the squared distance between the true and reconstructed pixels with intermediate values, we increase the importance of these pixels in the MSE loss calculation, encouraging the model to focus more on accurately reconstructing them. Furthermore, alternating between weighted and unweighted loss across epochs might help prevent the model from overfitting to particular pixel ranges. 

\begin{equation}
\text{L(\textbf{x})} = \frac{1}{N} \sum_{i} \left( (x_i - d(e(x_i)))^2 \cdot w_i \right)
\end{equation}
where
\begin{equation}
w_i = 
\begin{cases}  
2 & \text{if } 0 < x_i < 0.6 \text{ \& epoch $>$ $\alpha$ \& epoch \% $\beta$ $\neq$ 0} \\ 
1 & \text{otherwise}
\end{cases}
\end{equation}

Here $N$ is the total number of pixels,  $x_i$ is the input value of the $i^{th}$ pixel while $d(e(x_i))$ is its reconstructed value, the $w_i$ is the weight applied to each squared error based on the intensity of $x_i$ and the training epoch. While the optimal values of $\alpha$ and $\beta$ vary with the timescale of the images, we found that values around $\alpha \sim 100$ and $\beta \sim 4$ yield effective results across all timescales.

We train three ConvAE, one for each different timescale ``Long'', ``Medium'', and ``Short'' (see main manuscript). Initially, we attempted to use a single ConvAE for all timescales. However, the varying image sizes across the timescales required padding the ``Medium'' and ``Short'' timescale images with a significant number of zeros in order to match the ``Long'' timescale images size. This extensive padding adversely impacted the model's performance during the reconstruction process. Therefore, training three independent models, one for each timescale, proved to be a more effective approach. The ConvAE models are trained with 2361 events for 150 -- 250 epochs depending on the timescale (shorter timescales typically need more epochs as the images are more similar to each other and/or with less prominent features) and a batch size of 4. The main training parameters and the details on learning rate and activation functions used is provided in Table \ref{tab:nores_check}.

\renewcommand{\arraystretch}{1.2}

\begin{table}[htb]
\centering
\begin{ruledtabular}
\begin{tabular}{>{\hspace{0.2cm}}r<{\hspace{0.2cm}}>{\hspace{0.2cm}}c<{\hspace{0.2cm}}>{\hspace{0.2cm}}c<{\hspace{0.2cm}}>{\hspace{0.2cm}}c<{\hspace{0.2cm}}}
 & \textbf{Short TS} & \textbf{Medium TS} & \textbf{Long TS} \\
 \hline
\textbf{Epochs} & 250 & 250 & 200 \\
\textbf{Decaying LR ($\gamma$ - step)} & $8 \times 10^{-4}$ (0.9 - 10) & $8 \times 10^{-4}$ (0.9 - 10) & $1 \times 10^{-4}$ (0.9 - 10)\\
\hline
\textbf{Optimizer} & \multicolumn{3}{c}{Adam} \\
\textbf{Batch Size} & \multicolumn{3}{c}{4} \\
\textbf{Loss function} & \multicolumn{3}{c}{Mean Squared Error (MSE)} \\
\textbf{Activation functions} & \multicolumn{3}{c}{Leaky ReLU (Hidden layers) \& ReLU (Output layer)} \\
\end{tabular}
\caption{Autoencoder parameters used for each timescale (TS). Further details on the functions' descriptions are available in the \textit{pytorch} documentation.}
\label{tab:nores_check}
\end{ruledtabular}
\end{table}

\section{\label{app:D} ConvAE performance}

The general performance of the algorithm is evaluated both quantitatively by calculating the mean pixel-by-pixel absolute difference between the original input images and the ConvAE-reconstructed output images, and qualitatively by looking at the comparison between the output and input images for randomly selected sub-samples. 

\begin{figure*}
    \centering
    \includegraphics[width=1\textwidth]{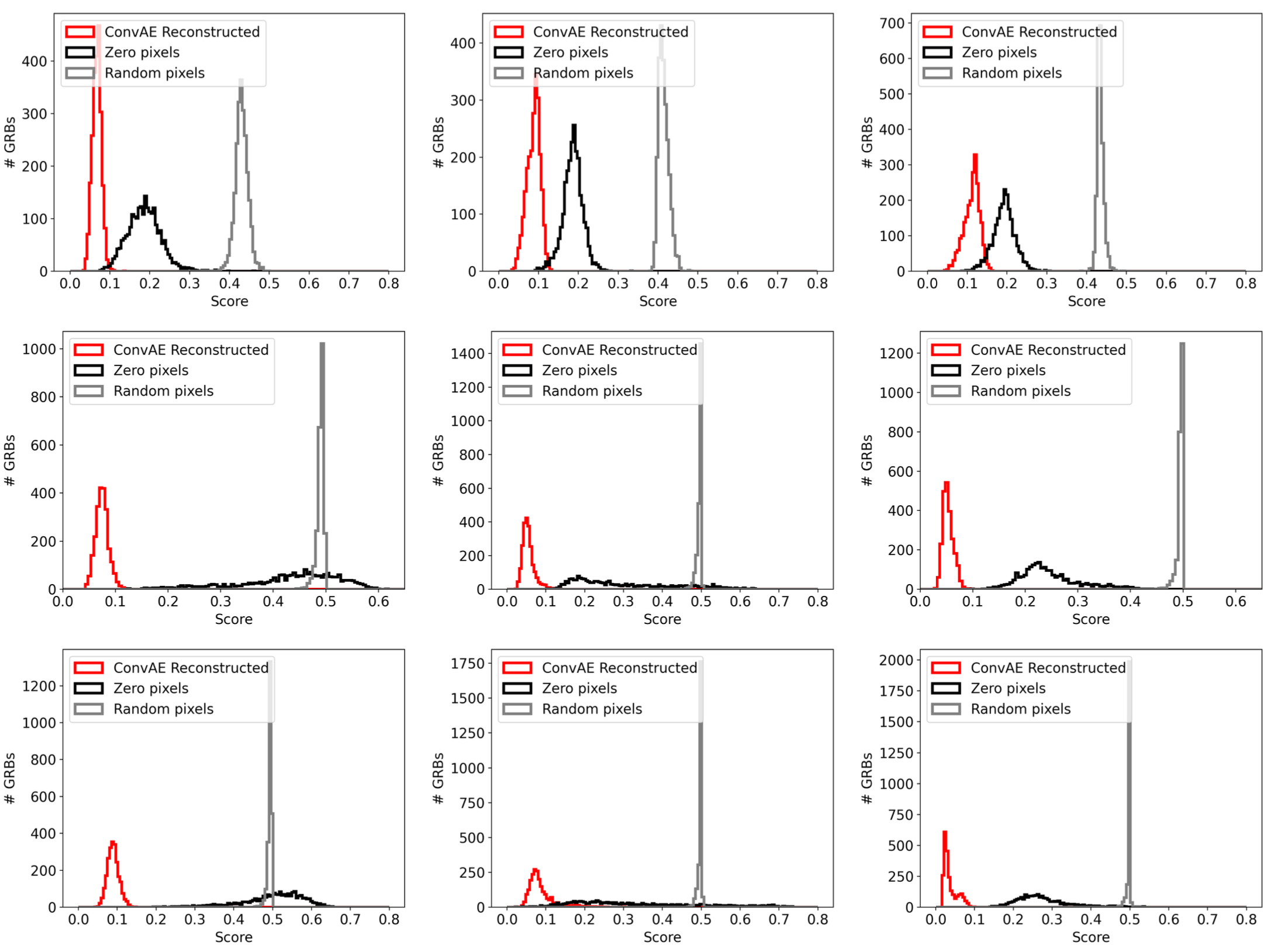}
    \caption{Top row: Mean difference (pixel-by-pixel) between the 4 long timescale input images and the ConvAE reconstructed ones (red histogram), denoted as \textit{Score}. The same quantity is also reported for a model which produces zero-pixels images (black histogram), and one which produces random [0,1] values for each pixel (grey histogram). (MinVal = 0). Middle row: Same as top row but for MinVal = 5. Bottom row: Same as top row but for MinVal = 10.}
    \label{fig:fig_merit_0}
\end{figure*}

The mean input-output difference is calculated for each GRB of the our training sample and for each timescale separately, treating the four images corresponding to different spectral assumptions as one single image. Because there are many zero-value pixels in the images (especially for medium- and short-timescales), they would disproportionately reduce the pixel-by-pixel difference and challenge the model comparison. Thus, the zero pixels reconstructed as zero are excluded from the performance evaluation of the model, whereas reconstructing zero pixels as non-zero affects negatively its performance. A representation of this figure of merit is shown in Figures~\ref{fig:fig_merit_0}, where the pixel-by-pixel mean difference is displayed for each GRB in the dataset and for all the models discussed in this document as a red histogram. The properties of this distribution, specifically the mean and variance, were used to compare the performance of the different models (corresponding to slightly different neural-network architecture and/or parametrization). To assess the effectiveness of the autoencoder in capturing the underlying structure of the data, we compare the reconstruction error (difference between the input and the autoencoder's output) to the difference between the input and an array of zeros (black histogram) or random values (gray histogram). Lower event scores relative to zero-scores  and random-scores suggests that the autoencoder is performing well at reconstructing the input data.

As per the visual assessment of the performance, we show the comparison of the Input vs Output (ConvAE-reconstructed) images for some of the GRBs used for the unblinding of the classification. Figures~\ref{fig:grb1},~\ref{fig:grb2},~\ref{fig:grb3},~\ref{fig:grb3},~\ref{fig:grb4},~\ref{fig:grb5},~\ref{fig:grb6},~\ref{fig:grb7}, show the True VS reconstructed image agreement for the case of MinVal=0.
Figures~\ref{fig:grb1a},~\ref{fig:grb2a},~\ref{fig:grb3a},~\ref{fig:grb3a},~\ref{fig:grb4a},~\ref{fig:grb5a},~\ref{fig:grb6a},~\ref{fig:grb7a}, refer to the MinVal=5 case. 
Figures~\ref{fig:grb1b},~\ref{fig:grb2b},~\ref{fig:grb3b},~\ref{fig:grb4b},~\ref{fig:grb5b},~\ref{fig:grb6b},~\ref{fig:grb7b}, refer to the MinVal=10 case.

 The red histograms demonstrate that the ConvAE models consistently produce low mean pixel-by-pixel differences, highlighting their ability to accurately reconstruct the input images. The qualitative comparisons further support this, showing a high degree of visual similarity between the reconstructed and true images for the GRBs used for the unblinding. This agreement across various metrics underscores the robustness and accuracy of the ConvAE approach in image reconstruction tasks, validating its effectiveness in capturing and replicating the details of the waterfall plots.

\section{\label{app:E}Dimensionality reduction}
The 10-dimensional latent spaces obtained from the different ConvAEs, are concatenated to generate a 30-dimensional representation of our sample of GRBs. Prior to the concatenation, the parameters of the latent space is normalized in order to have values between 0 and 1:
$$LS_i = \frac{LS_i - {\rm min}(LS_i)}{{\rm max}(LS_i) - {\rm min}(LS_i)}$$
We use the Uniform Manifold Approximation and Projection for Dimension Reduction (UMAP, \citep{UMAP}) to further reduce the dimensions of the combined 30-Dimension latent space. Most UMAP parameters are kept at their default values since altering them does not significantly impact the final distribution. The parameters that are adjusted are done so according to the dataset’s characteristics (known prior to unblinding the dataset for inference and validation). For example, ‘n\_neighbors’ is set to values commonly used in similar datasets, ‘min\_dist’ is set to 0 to avoid restricting the final distance between points, and ‘learning\_rate’ and ‘n\_epochs’ are adjusted to ensure the final distribution is as stable as possible across different UMAP runs. We set the UMAP parameters to the following values:
\begin{itemize}
    \item \tt{n\_neighbors = 30}
    \item \tt{min\_dist = 0.}
    \item \tt{n\_components = 3 (or 2)}
    \item \tt{metric = `euclidean'}
    \item \tt{local\_connectivity = 0.5}
    \item \tt{n\_epochs = 1000}
    \item \tt{learning\_rate=0.001}
\end{itemize}

\subsection{The case of the long mergers}
GRB~230307A \citep{levan2024heavy} and GRB~211211A \citep{rastinejad2022kilonova} are two \textit{long} GRBs ($\sim$ 100\,s) where the prompt emission is followed by a kilonova-like thermal transient, indicating a merger origin. One would expect that the distinctive signature informing us that these two GRBs are mergers resides in the \textit{short} timescale. In fact, looking at the 3D embeddings in Figure \ref{fig:latent_long_mergers} obtained through dimensionality reduction of the three different latent spaces, we can notice how the two long merger events are co-spatial only for the \textit{short} timescale. The fact that the two GRBs fall close to each other in the final embedding that combines the three latent spaces suggests that the \textit{short} timescales drives the UMAP's dimensionality reduction more than the \textit{medium} and \textit{long} ones for these events. To investigate this intuition, we look at the distributions of the values of the 10-dimensional latent spaces, which are reported in the left panel of Figure \ref{fig:LatentDistrib}. Note how different the non-normalized latent space distributions are, not only in shape but also in the range of absolute values. We suspect that this is due to the greater uniformity in the \textit{short} timescale images, which offers typically fewer sharp features than the longer timescales. Such a big difference in absolute scales among the three latent spaces cannot be only attributed to the initial differences in the input images values (which we remind are normalized to the global maximum across the timescales for each GRB), but might be also related to the different performance of the ConvAEs, trained separately for each timescale. On the other hand, for the dimensionality reduction algorithm, such discrepancies in absolute values is not ideal, because it would result in an non-physical unbalance between the importance of the timescales. Therefore, before merging the three 10-D latent spaces into the 30-D one, we normalize their parameters to be between 0 and 1, as shown in the right panel of Figure \ref{fig:LatentDistrib}. We also tried to normalize each 10-D latent space parameters to their maximum (without forcing the values to be positive) with no changes in the conclusions.

\begin{figure*}
    \centering
    \includegraphics[width=\textwidth]{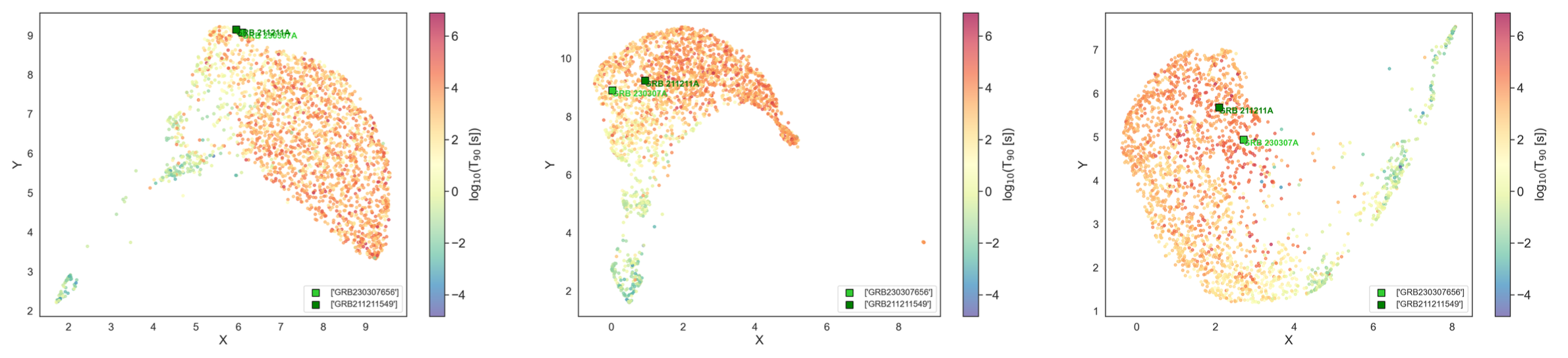}
    \caption{3D embeddings for \textit{short} (left), \textit{medium} (middle), and \textit{long} (right) timescales (for the case of minval=10). We mark the long mergers, GRB~230307A \citep{levan2024heavy} and GRB~211211A \citep{rastinejad2022kilonova}. }
    \label{fig:3latents}
\end{figure*}

\begin{figure*}
    \centering
    \includegraphics[width=\textwidth]{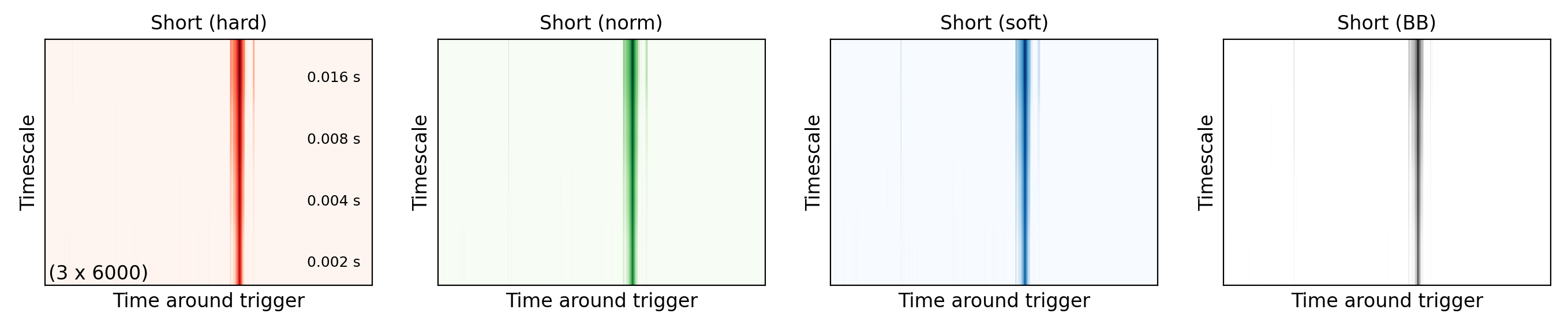}
    \includegraphics[width=\textwidth]{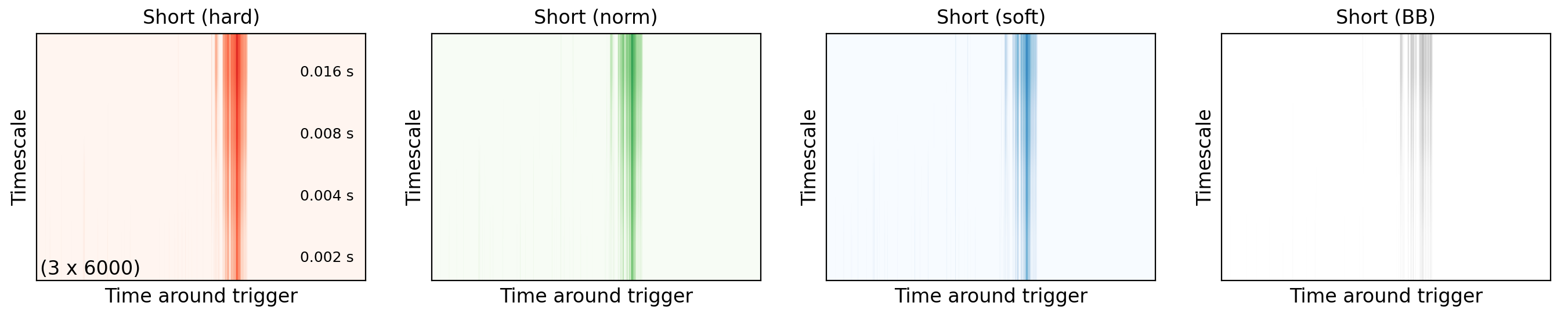}
    \caption{Waterfall plots of \textit{short} timescales for GRB~211211A \citep{rastinejad2022kilonova} (top) and GRB~230307A \citep{levan2024heavy} (bottom).}
    \label{fig:latent_long_mergers}
\end{figure*}

\begin{figure*}
    \centering
    \includegraphics[width=0.46\textwidth]{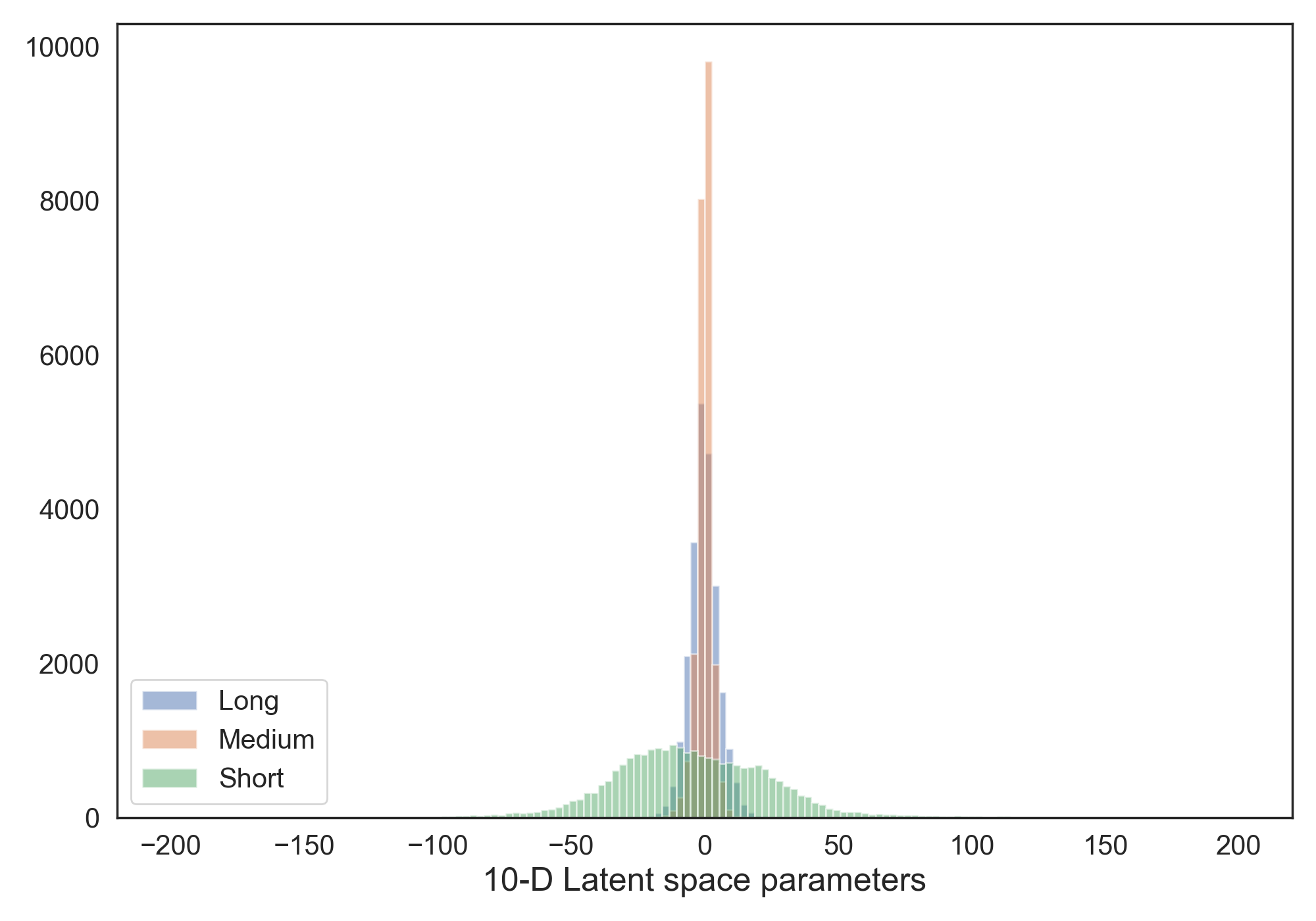}~~~
    \includegraphics[width=0.46\textwidth]{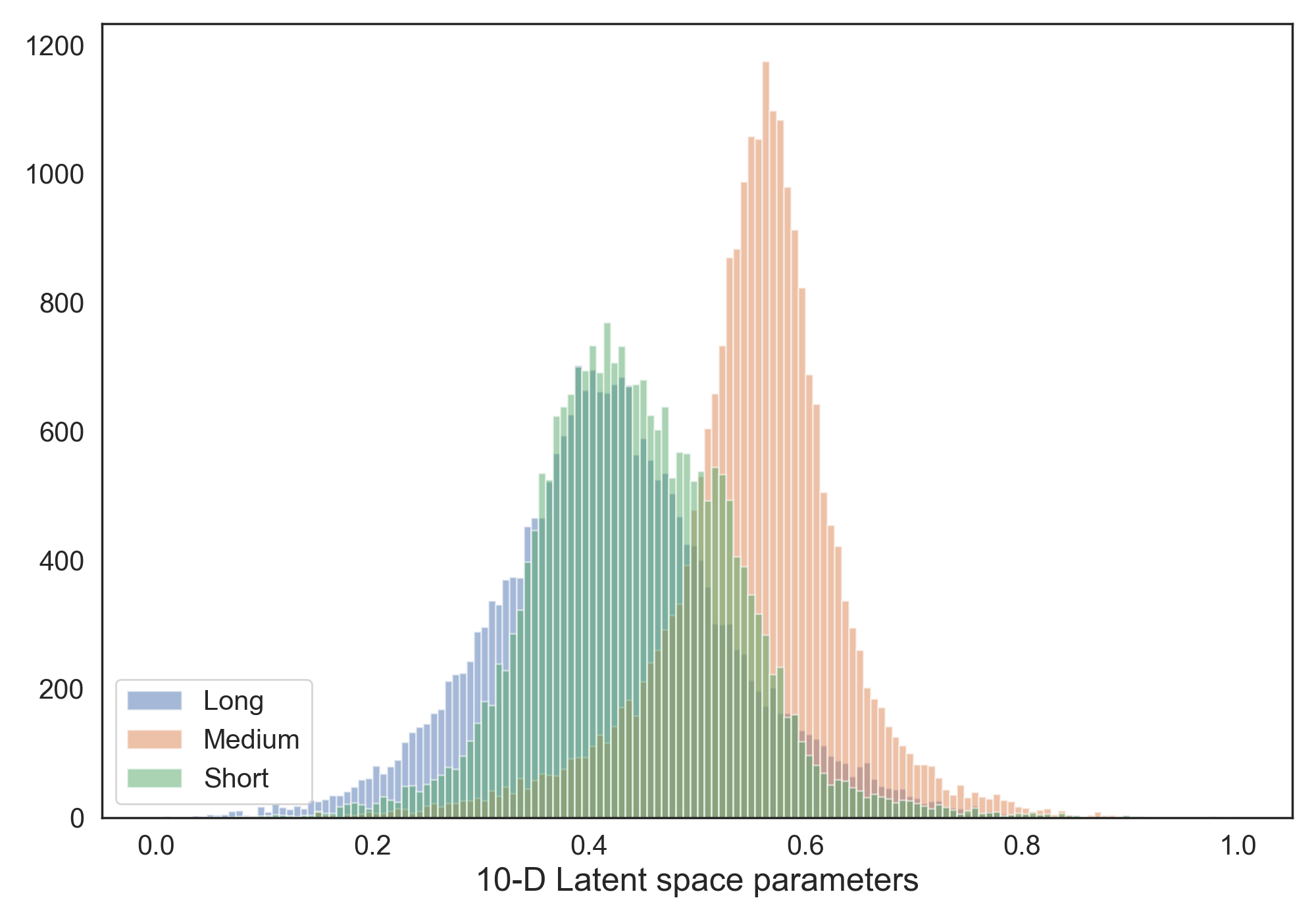}
    \caption{Distribution of the (flattened) 10-dimensional Latent spaces' non-normalized (left) and normalized (right) parameters for \textit{short}, \textit{medium}, and \textit{long} timescales. }
    \label{fig:LatentDistrib}
\end{figure*}

\begin{figure*}
    \centering
    \includegraphics[width=\textwidth]{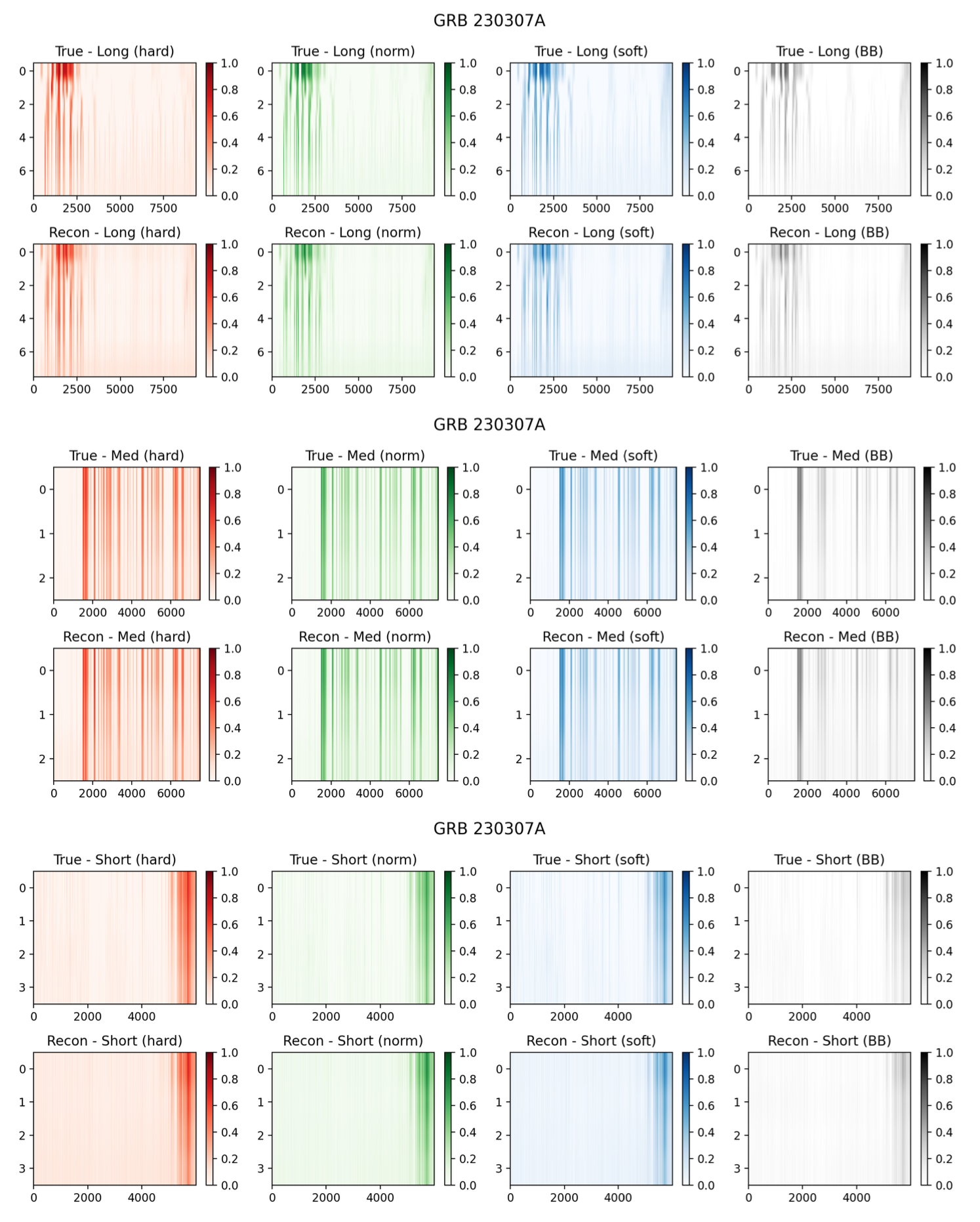}
    \caption{Reconstructed VS True waterfalls for GRB~230307A (MinVal = 0)}
    \label{fig:grb1}
\end{figure*}

\begin{figure*}
    \centering
    \includegraphics[width=\textwidth]{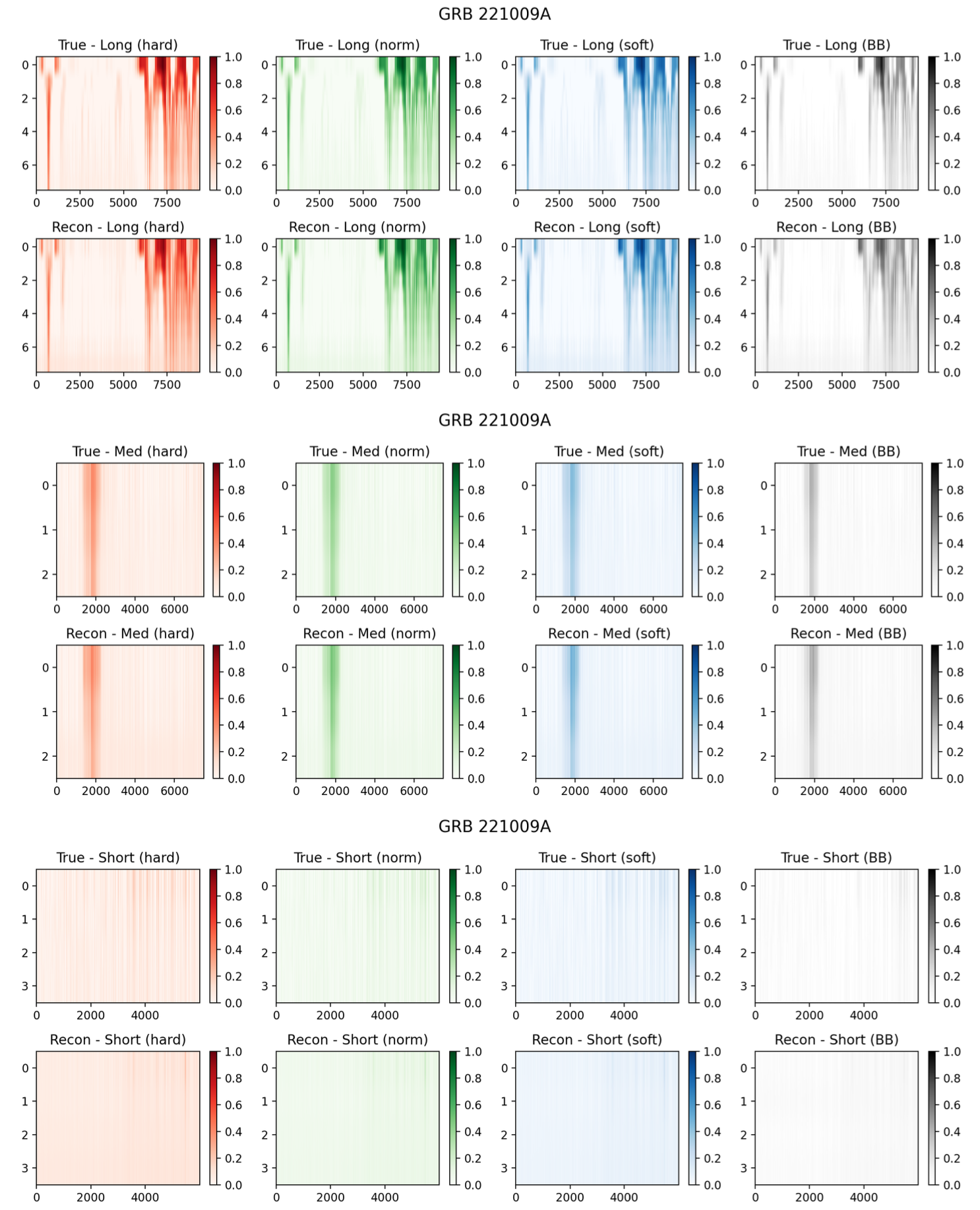}
    \caption{Reconstructed VS True waterfalls for GRB~221009A (MinVal = 0)}
    \label{fig:grb2}
\end{figure*}

\begin{figure*}
    \centering
    \includegraphics[width=\textwidth]{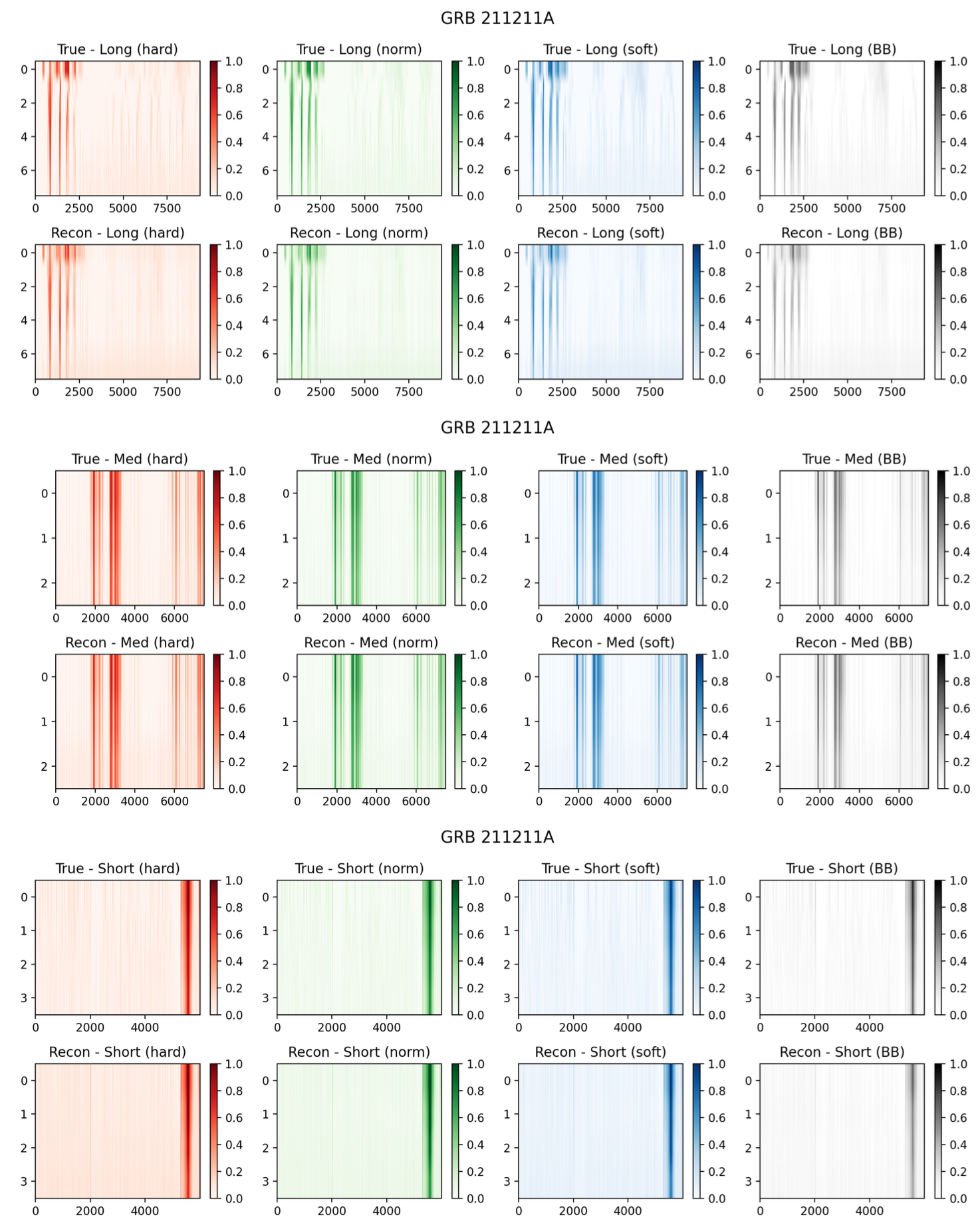}
    \caption{Reconstructed VS True waterfalls for GRB~211211A (MinVal = 0)}
    \label{fig:grb3}
\end{figure*}

\begin{figure*}
    \centering
    \includegraphics[width=\textwidth]{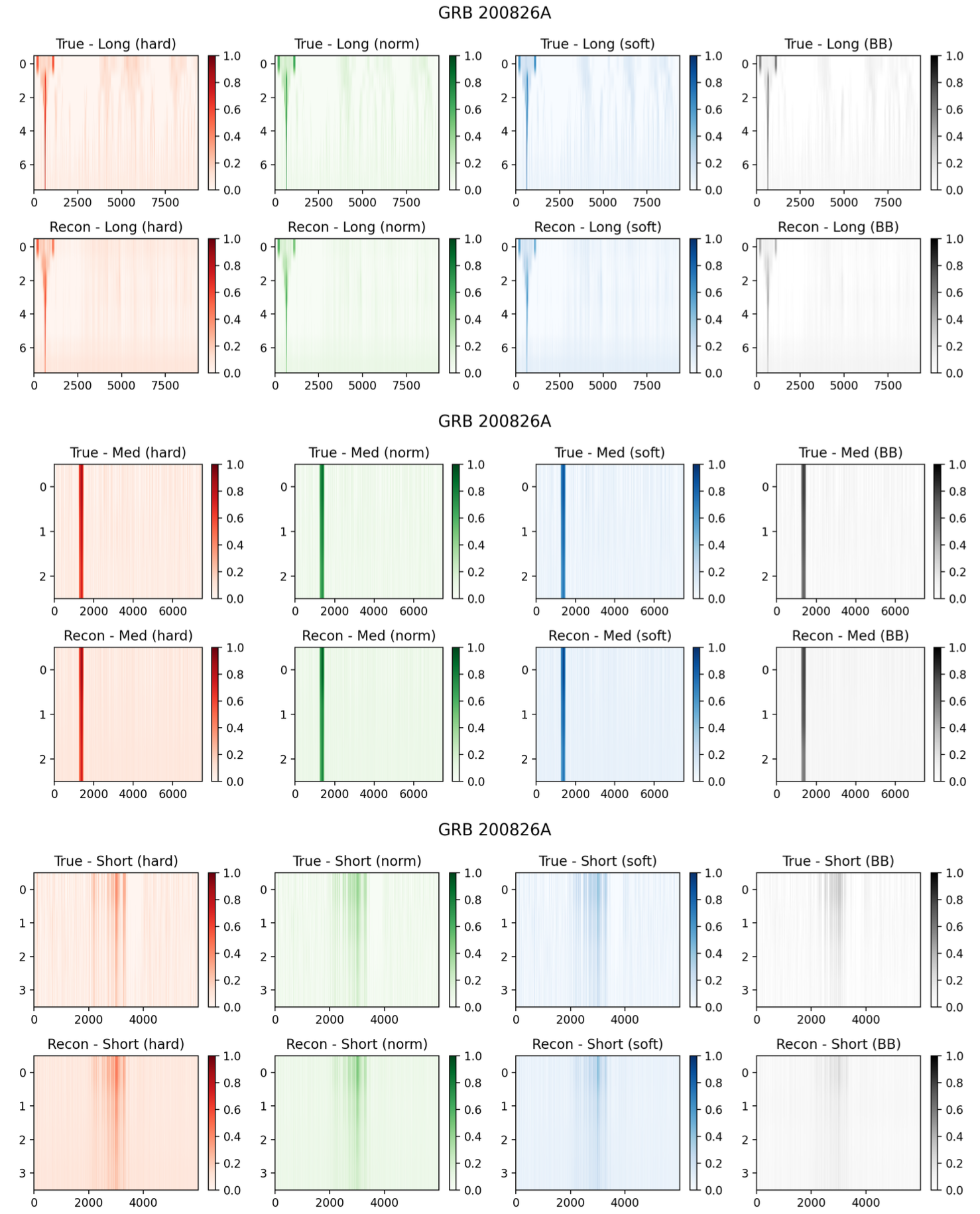}
    \caption{Reconstructed VS True waterfalls for GRB~200826A (MinVal = 0)}
    \label{fig:grb4}
\end{figure*}

\begin{figure*}
    \centering
    \includegraphics[width=\textwidth]{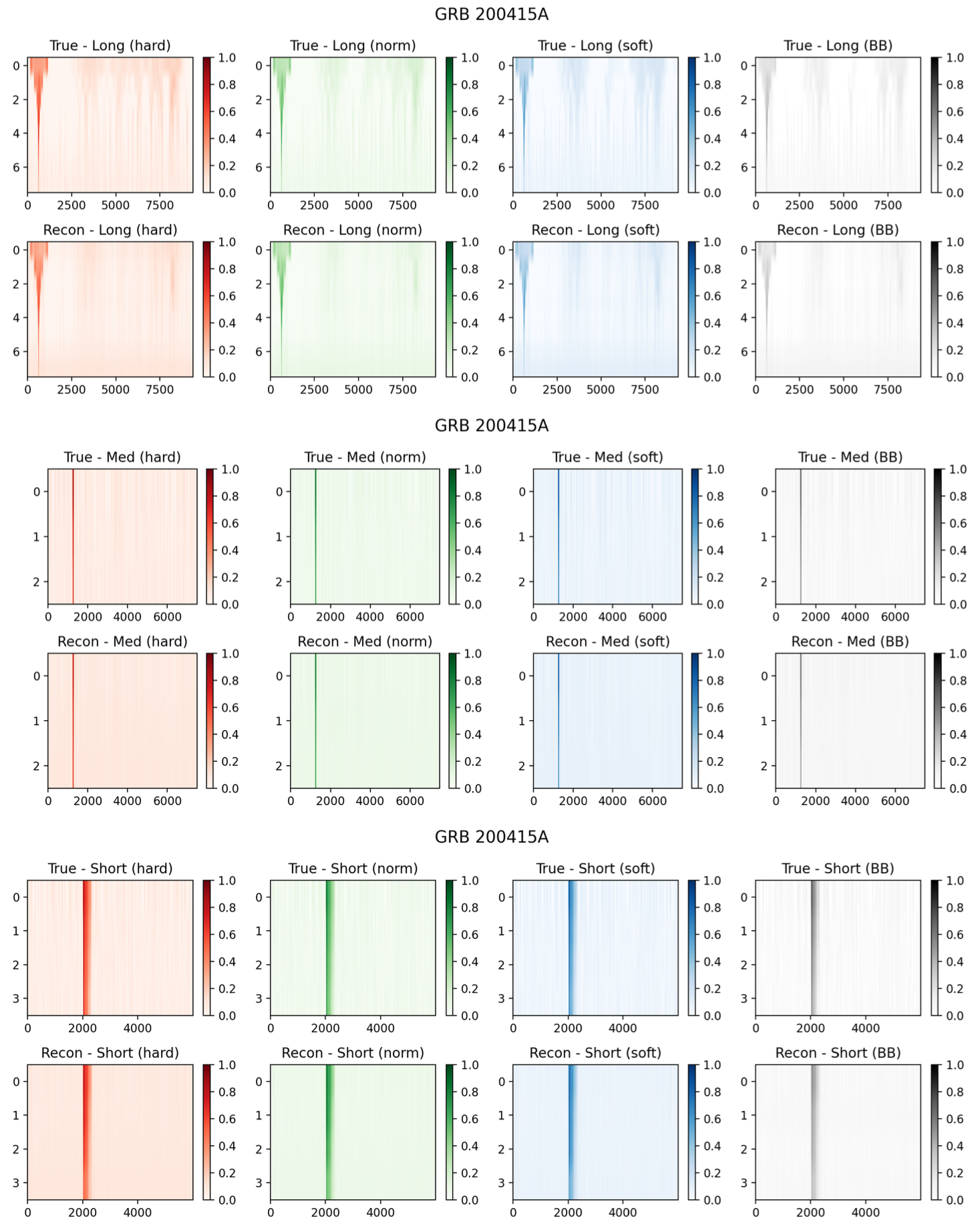}
    \caption{Reconstructed VS True waterfalls for GRB~200415A (MinVal = 0)}
    \label{fig:grb5}
\end{figure*}

\begin{figure*}
    \centering
    \includegraphics[width=\textwidth]{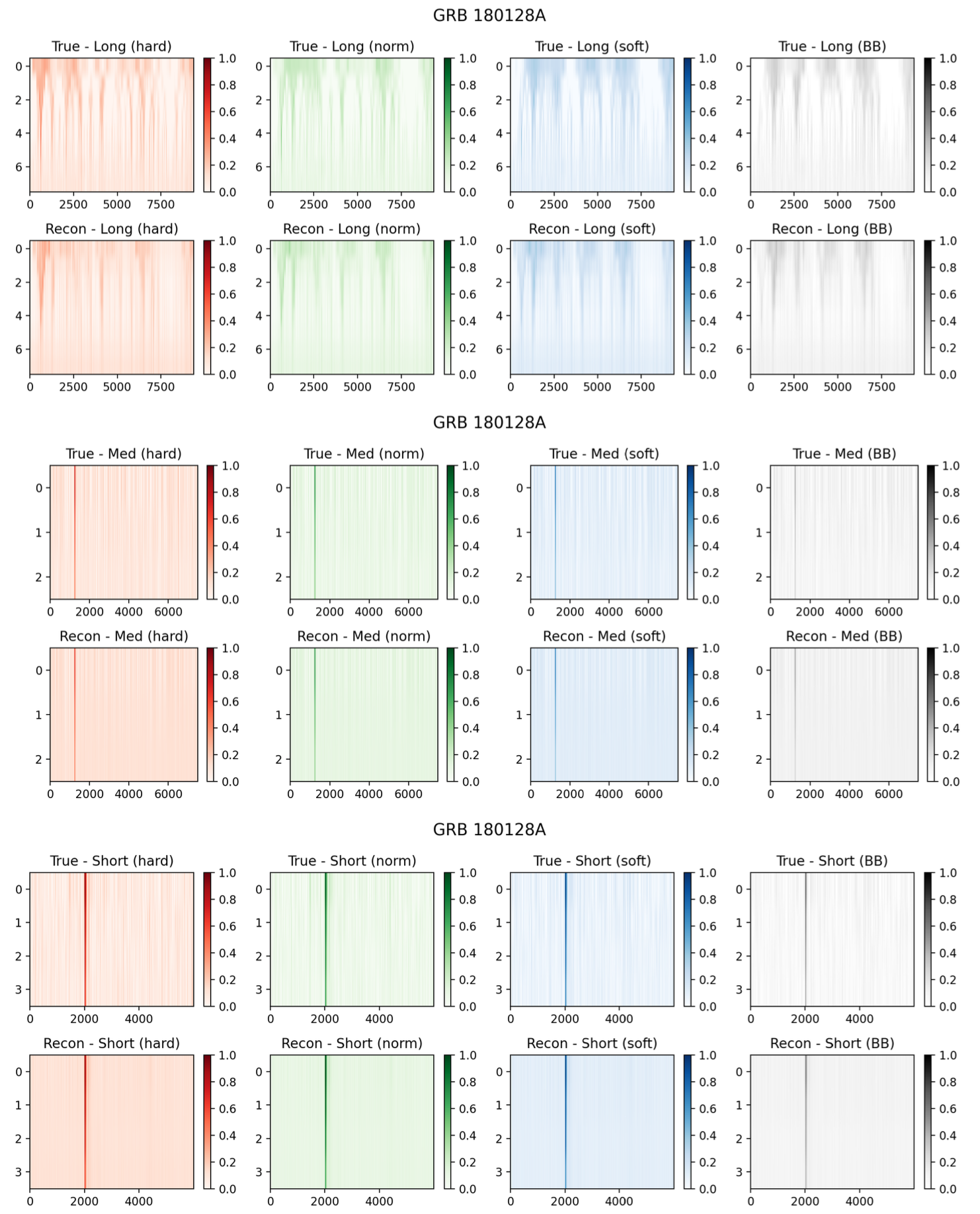}
    \caption{Reconstructed VS True waterfalls for GRB~180128A (MinVal = 0)}
    \label{fig:grb6}
\end{figure*}

\begin{figure*}
    \centering
    \includegraphics[width=\textwidth]{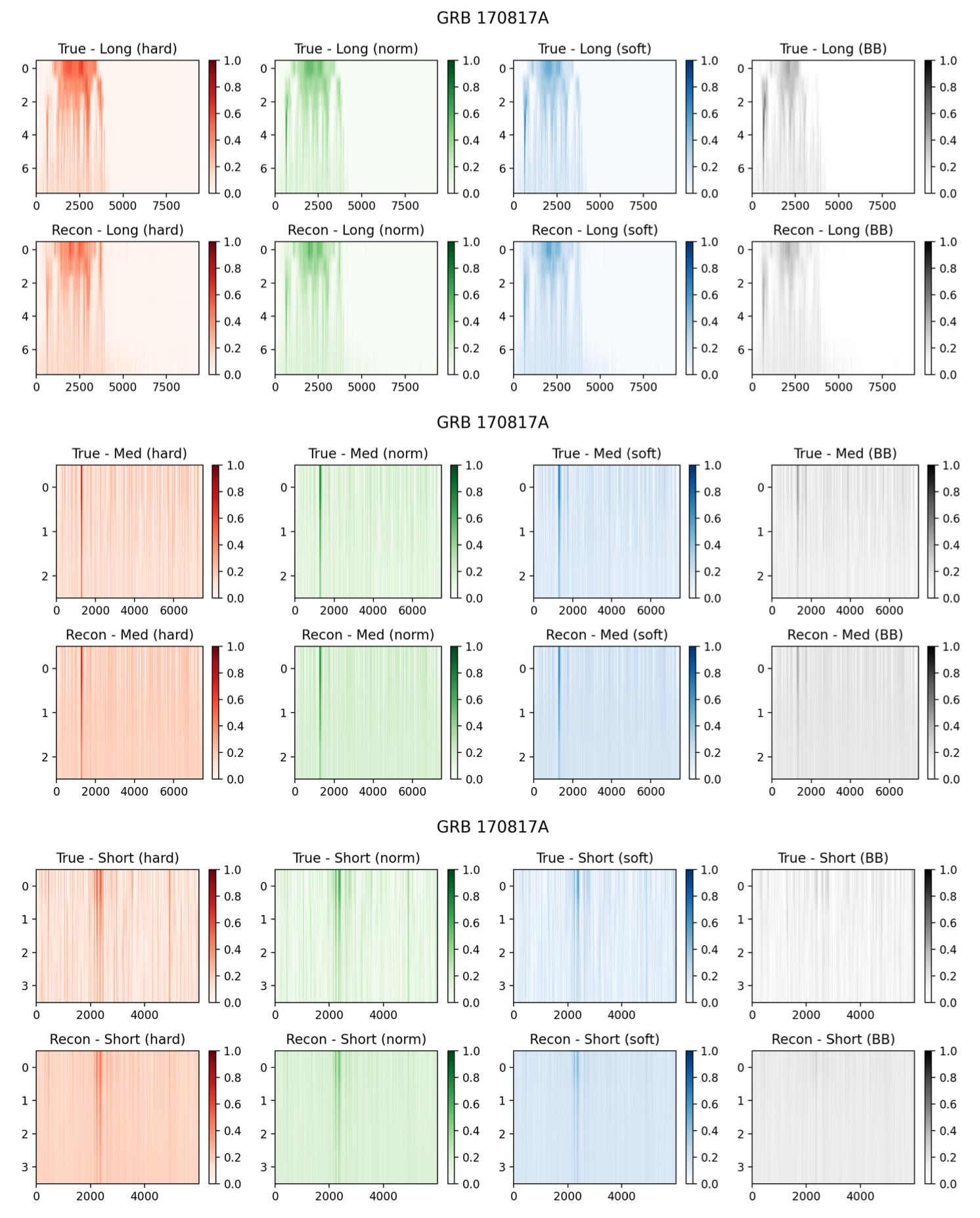}
    \caption{Reconstructed VS True waterfalls for GRB~170817A (MinVal = 0)}
    \label{fig:grb7}
\end{figure*}


\begin{figure*}
    \centering
    \includegraphics[width=\textwidth]{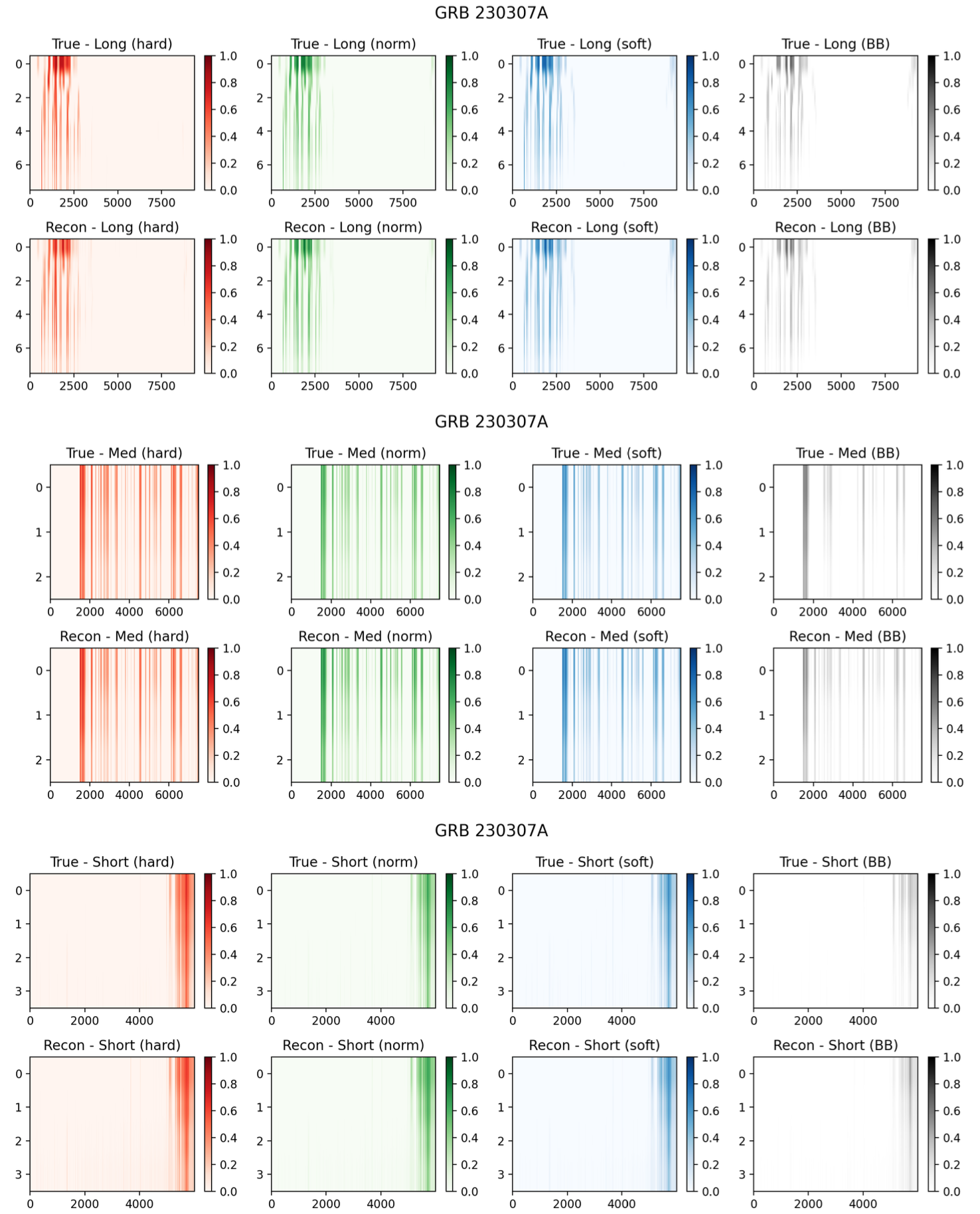}
    \caption{Reconstructed VS True waterfalls for GRB~230307A (MinVal = 5)}
    \label{fig:grb1a}
\end{figure*}

\begin{figure*}
    \centering
    \includegraphics[width=\textwidth]{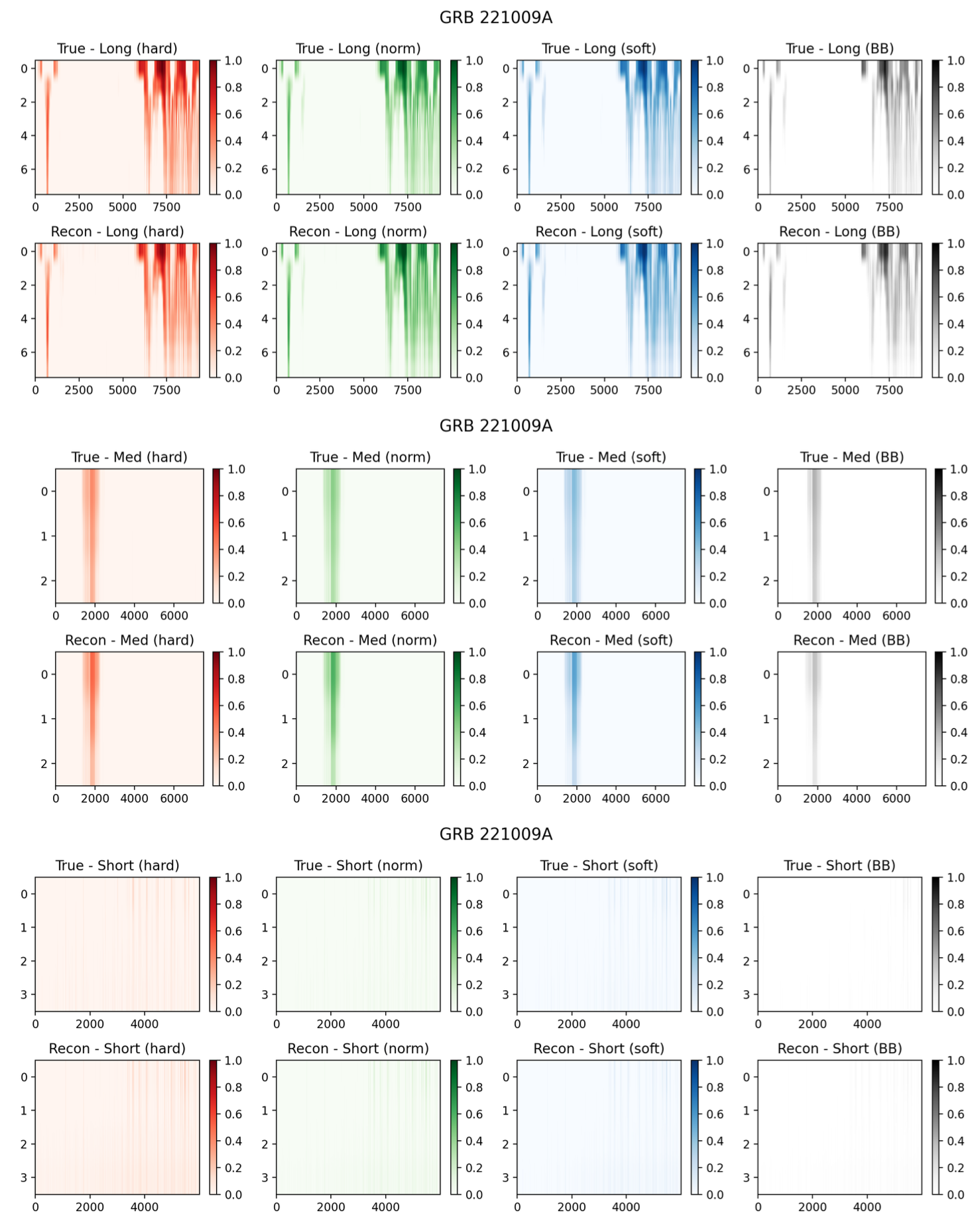}
    \caption{Reconstructed VS True waterfalls for GRB~221009A (MinVal = 5)}
    \label{fig:grb2a}
\end{figure*}

\begin{figure*}
    \centering
    \includegraphics[width=\textwidth]{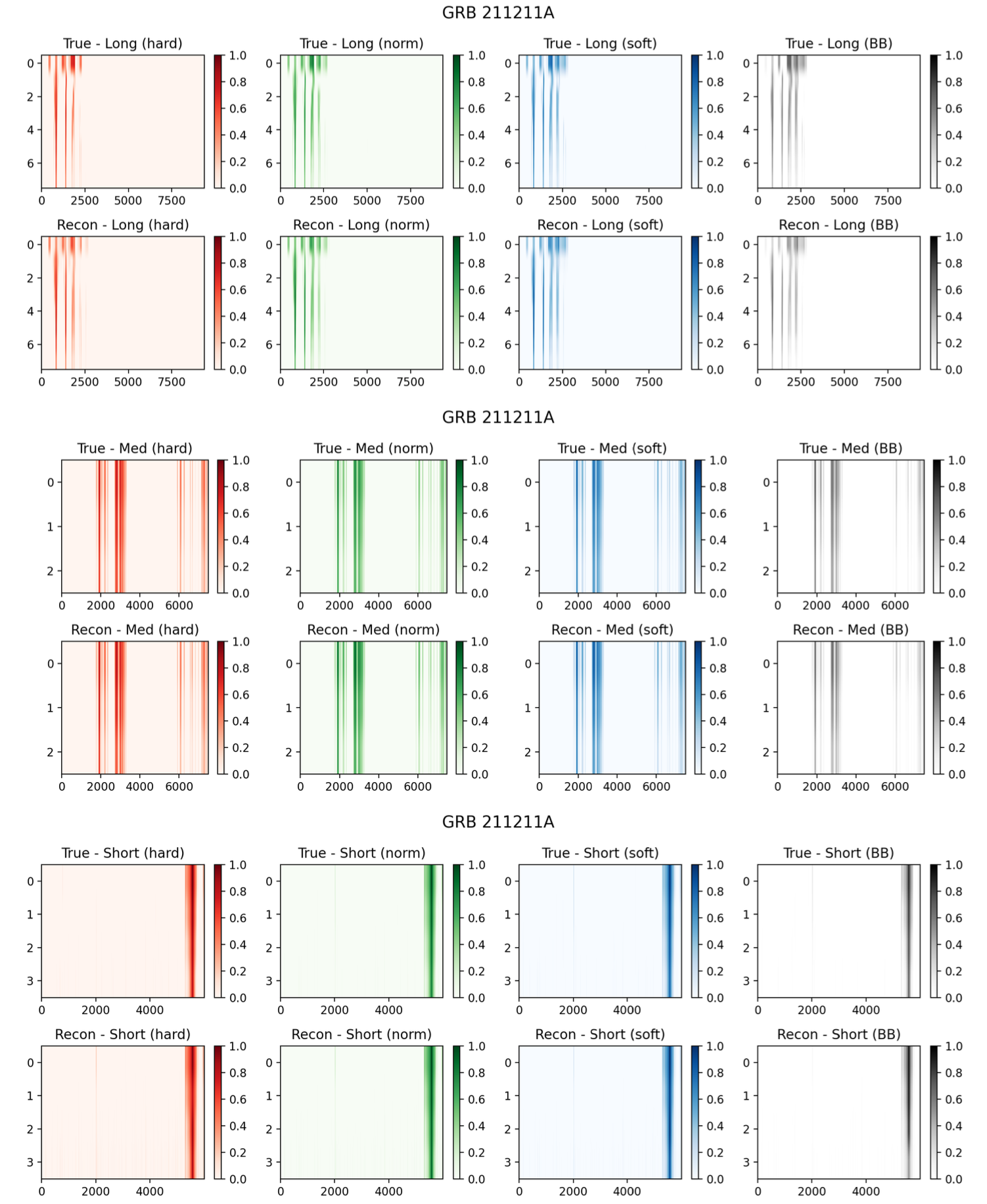}
    \caption{Reconstructed VS True waterfalls for GRB~211211A (MinVal = 5)}
    \label{fig:grb3a}
\end{figure*}

\begin{figure*}
    \centering
    \includegraphics[width=\textwidth]{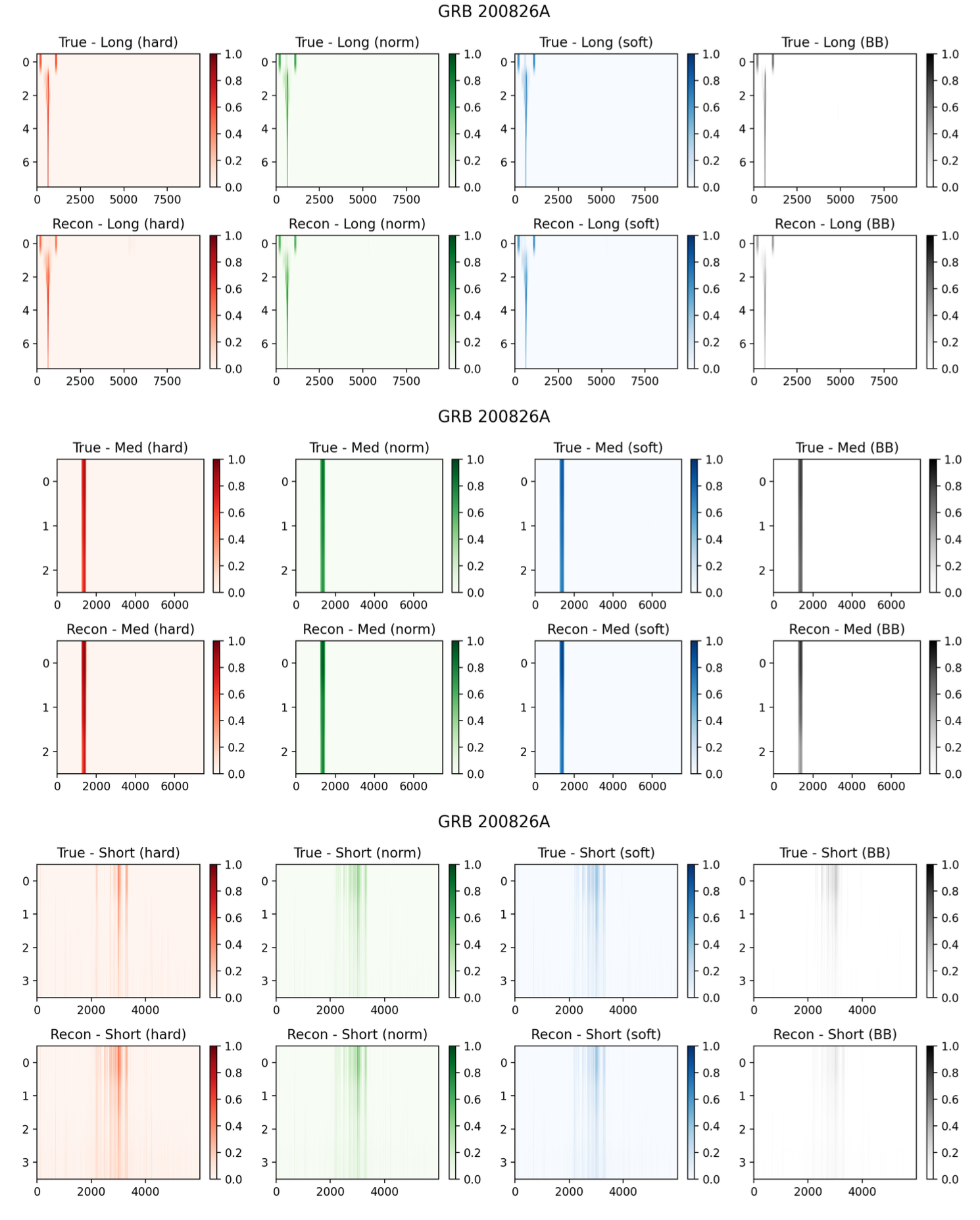}
    \caption{Reconstructed VS True waterfalls for GRB~200826A (MinVal = 5)}
    \label{fig:grb4a}
\end{figure*}

\begin{figure*}
    \centering
    \includegraphics[width=\textwidth]{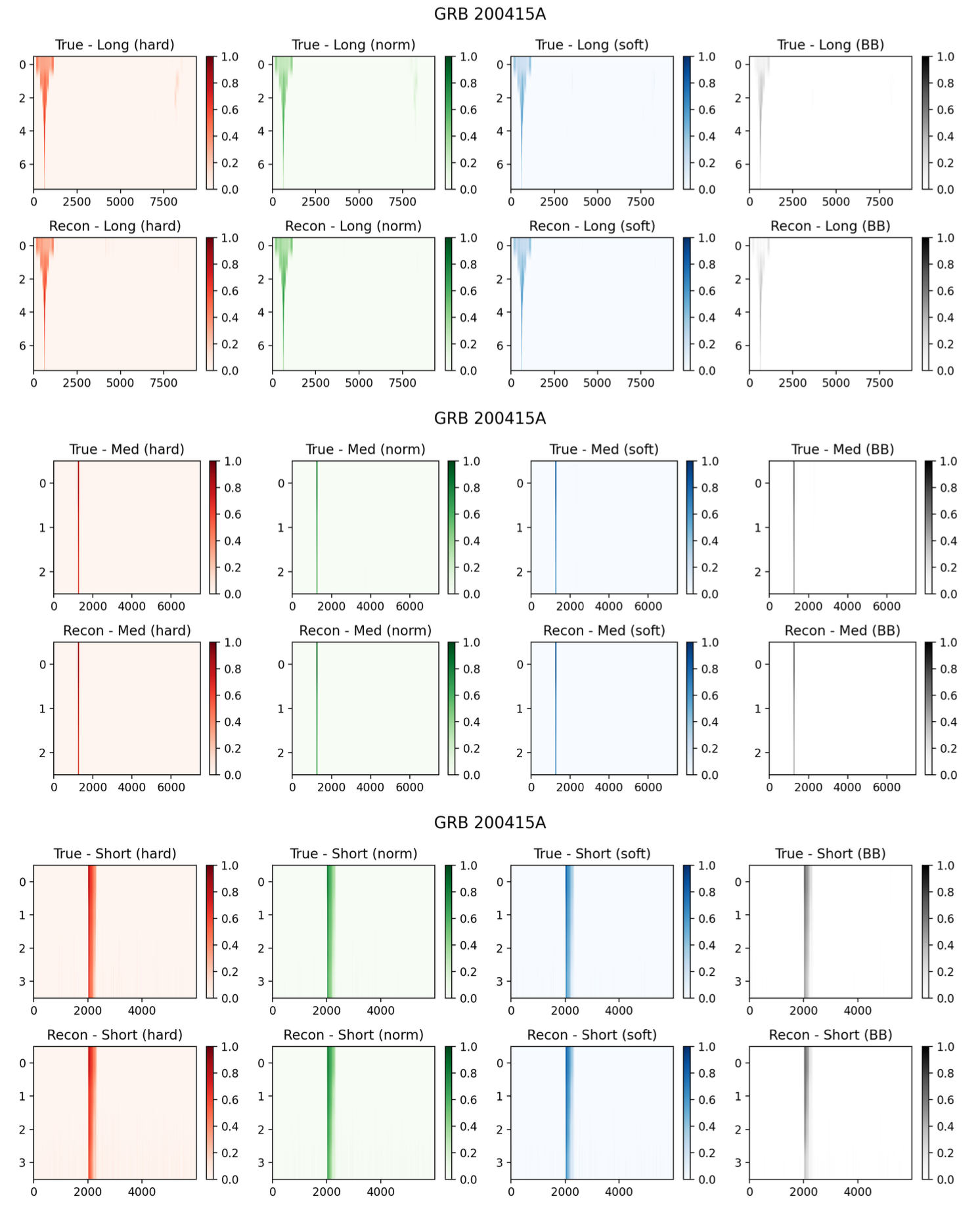}
    \caption{Reconstructed VS True waterfalls for GRB~200415A (MinVal = 5)}
    \label{fig:grb5a}
\end{figure*}

\begin{figure*}
    \centering
    \includegraphics[width=\textwidth]{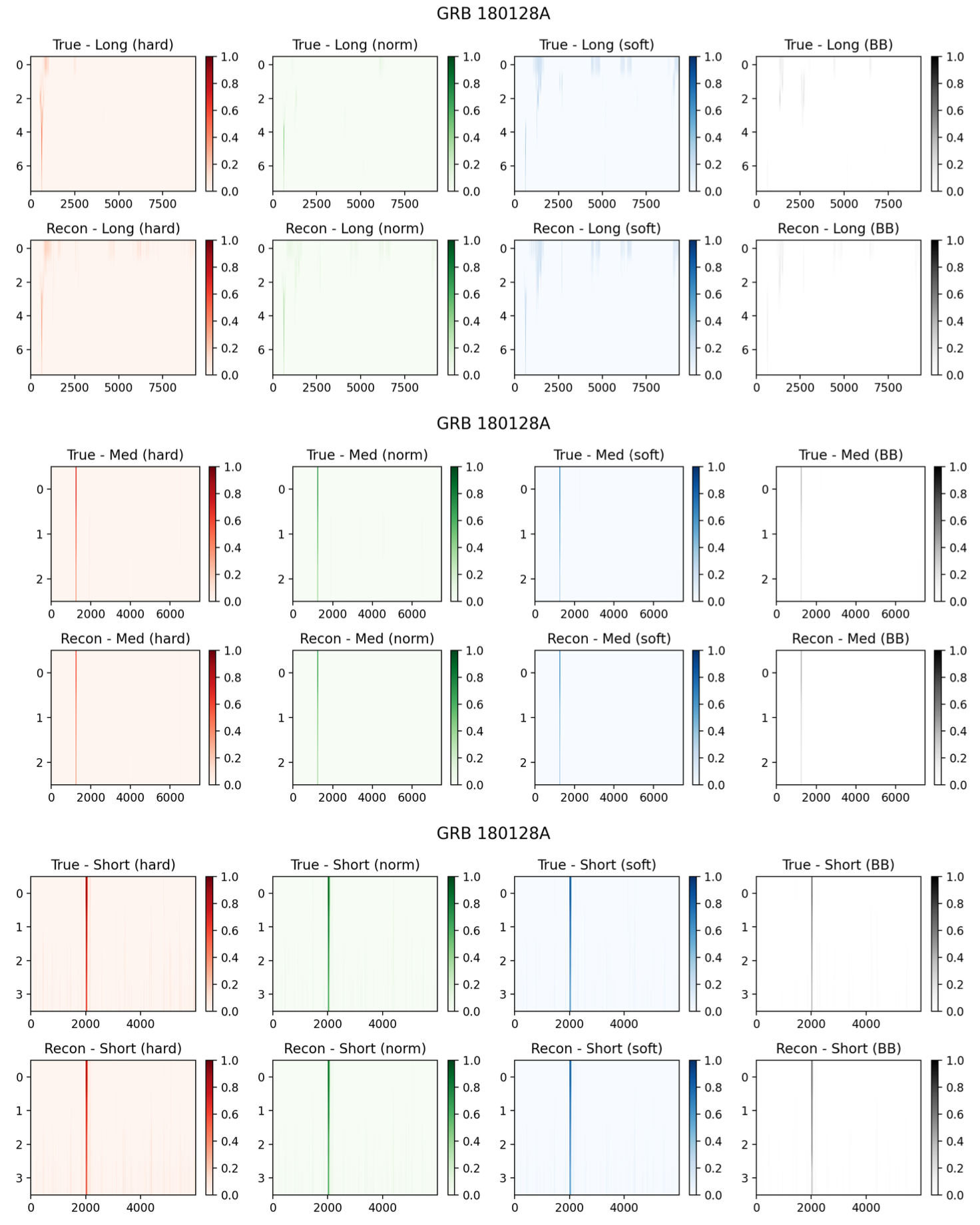}
    \caption{Reconstructed VS True waterfalls for GRB~180128A (MinVal = 5)}
    \label{fig:grb6a}
\end{figure*}

\begin{figure*}
    \centering
    \includegraphics[width=\textwidth]{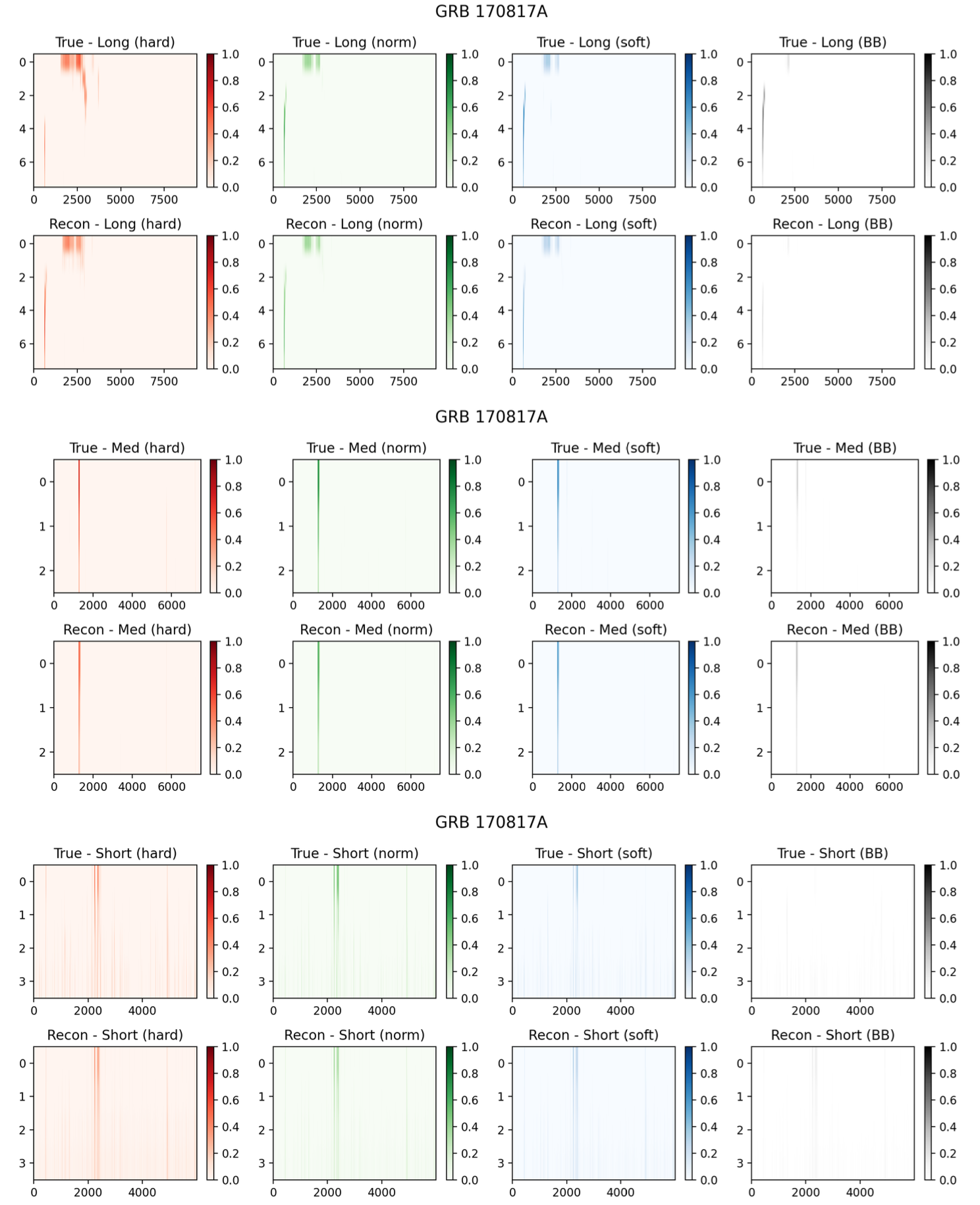}
    \caption{Reconstructed VS True waterfalls for GRB~170817A (MinVal = 5)}
    \label{fig:grb7a}
\end{figure*}


\begin{figure*}
    \centering
    \includegraphics[width=\textwidth]{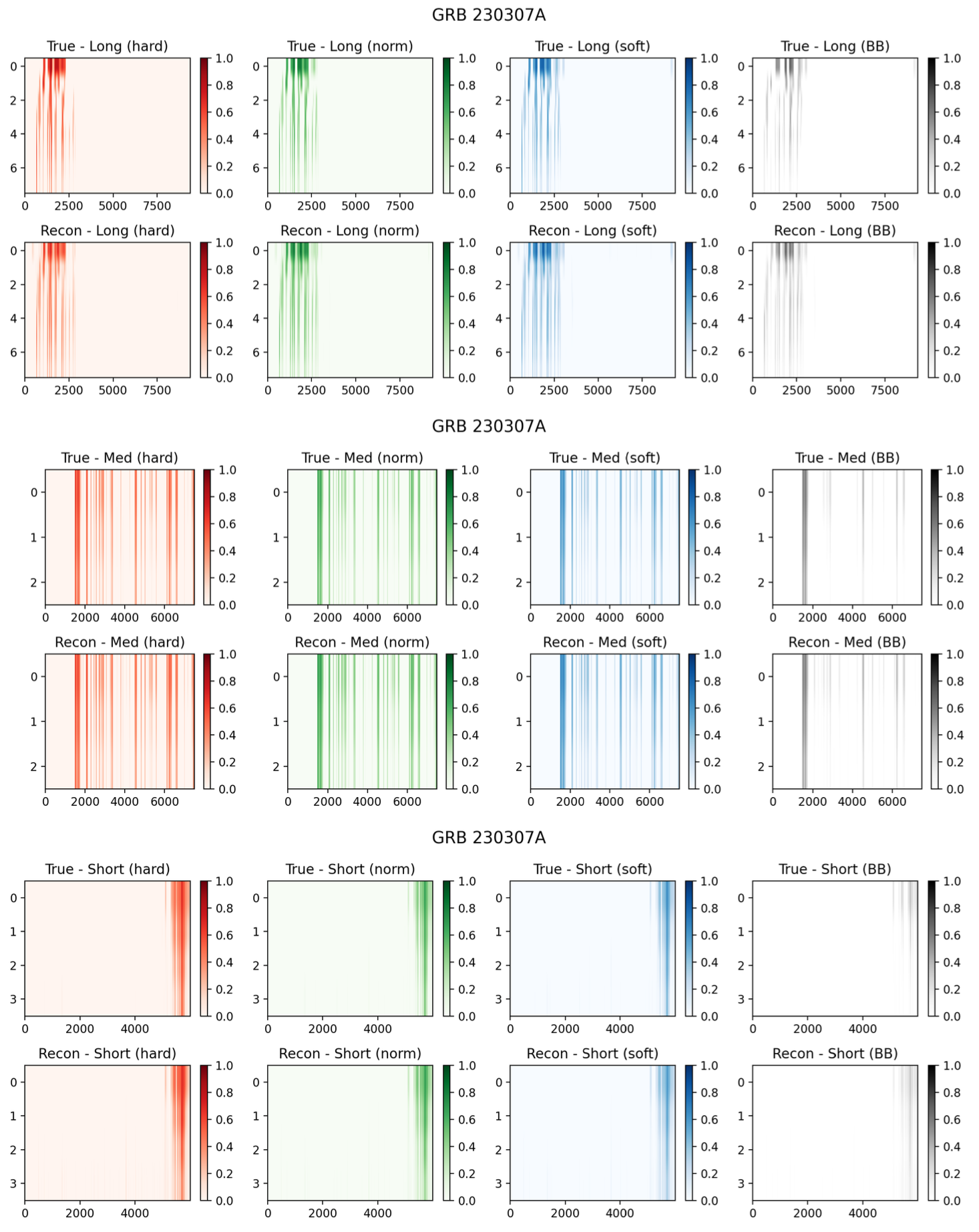}
    \caption{Reconstructed VS True waterfalls for GRB~230307A (MinVal = 10)}
    \label{fig:grb1b}
\end{figure*}

\begin{figure*}
    \centering
    \includegraphics[width=\textwidth]{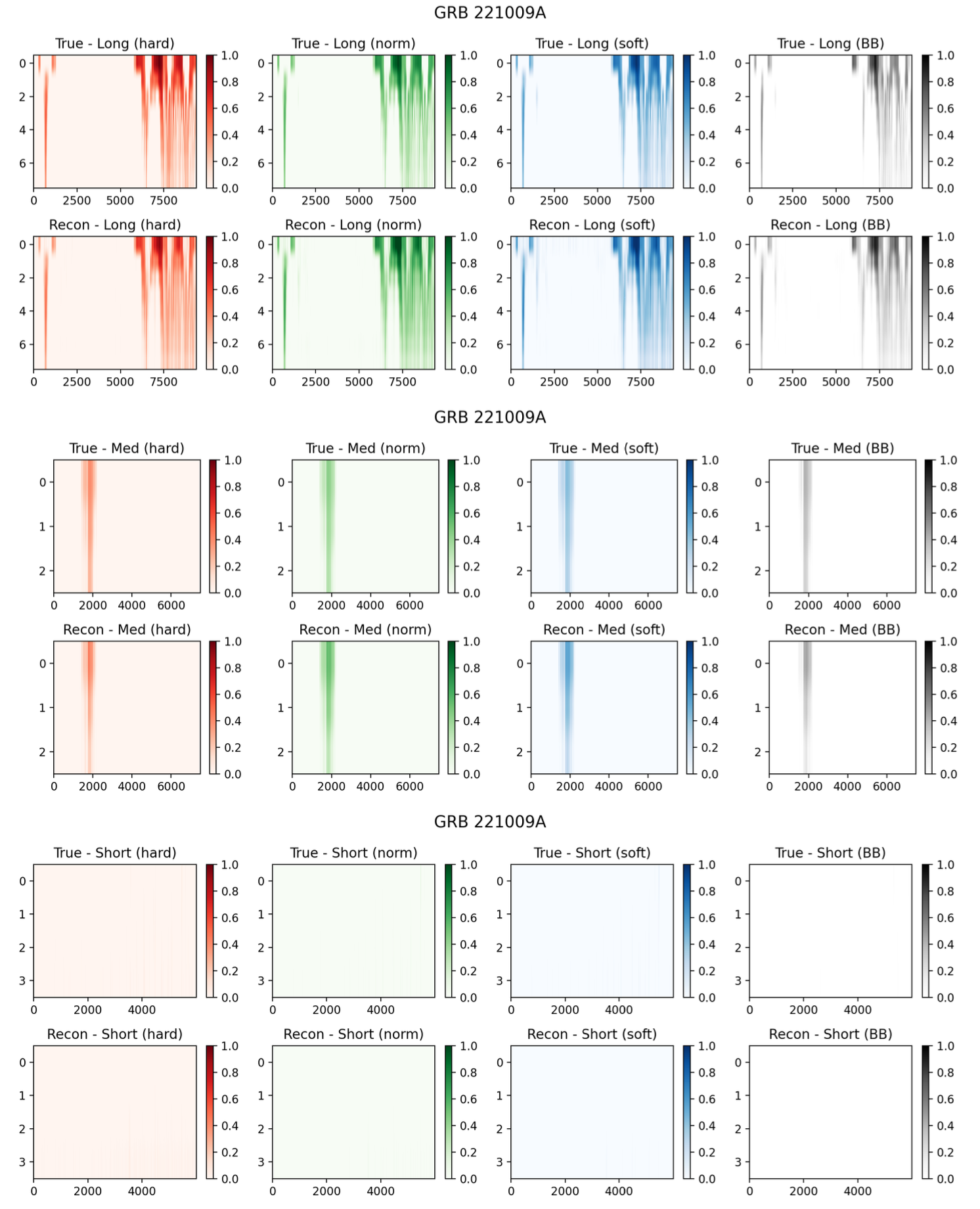}
    \caption{Reconstructed VS True waterfalls for GRB~221009A (MinVal = 10)}
    \label{fig:grb2b}
\end{figure*}

\begin{figure*}
    \centering
    \includegraphics[width=\textwidth]{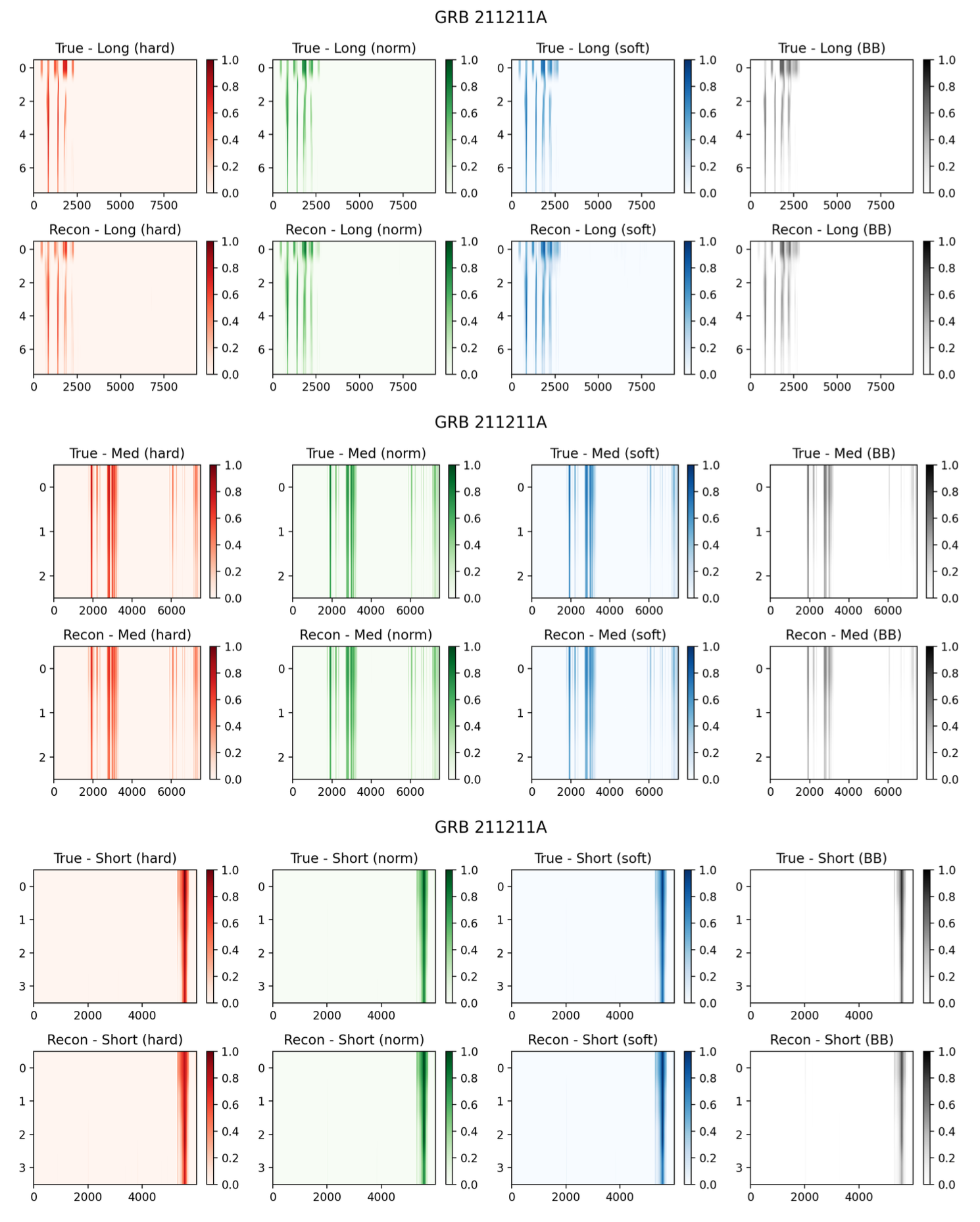}
    \caption{Reconstructed VS True waterfalls for GRB~211211A (MinVal = 10)}
    \label{fig:grb3b}
\end{figure*}

\begin{figure*}
    \centering
    \includegraphics[width=\textwidth]{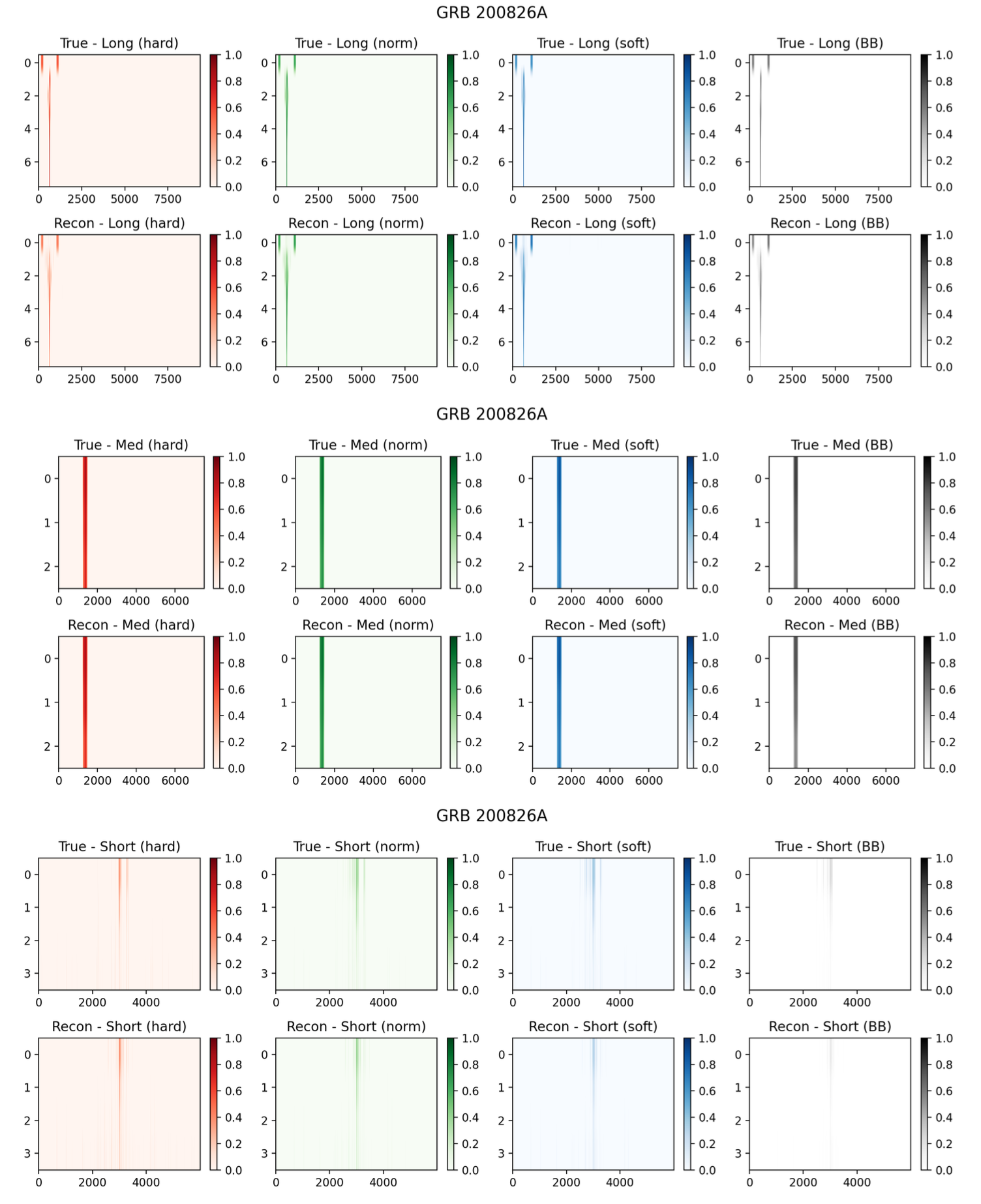}
    \caption{Reconstructed VS True waterfalls for GRB~200826A (MinVal = 10)}
    \label{fig:grb4b}
\end{figure*}

\begin{figure*}
    \centering
    \includegraphics[width=\textwidth]{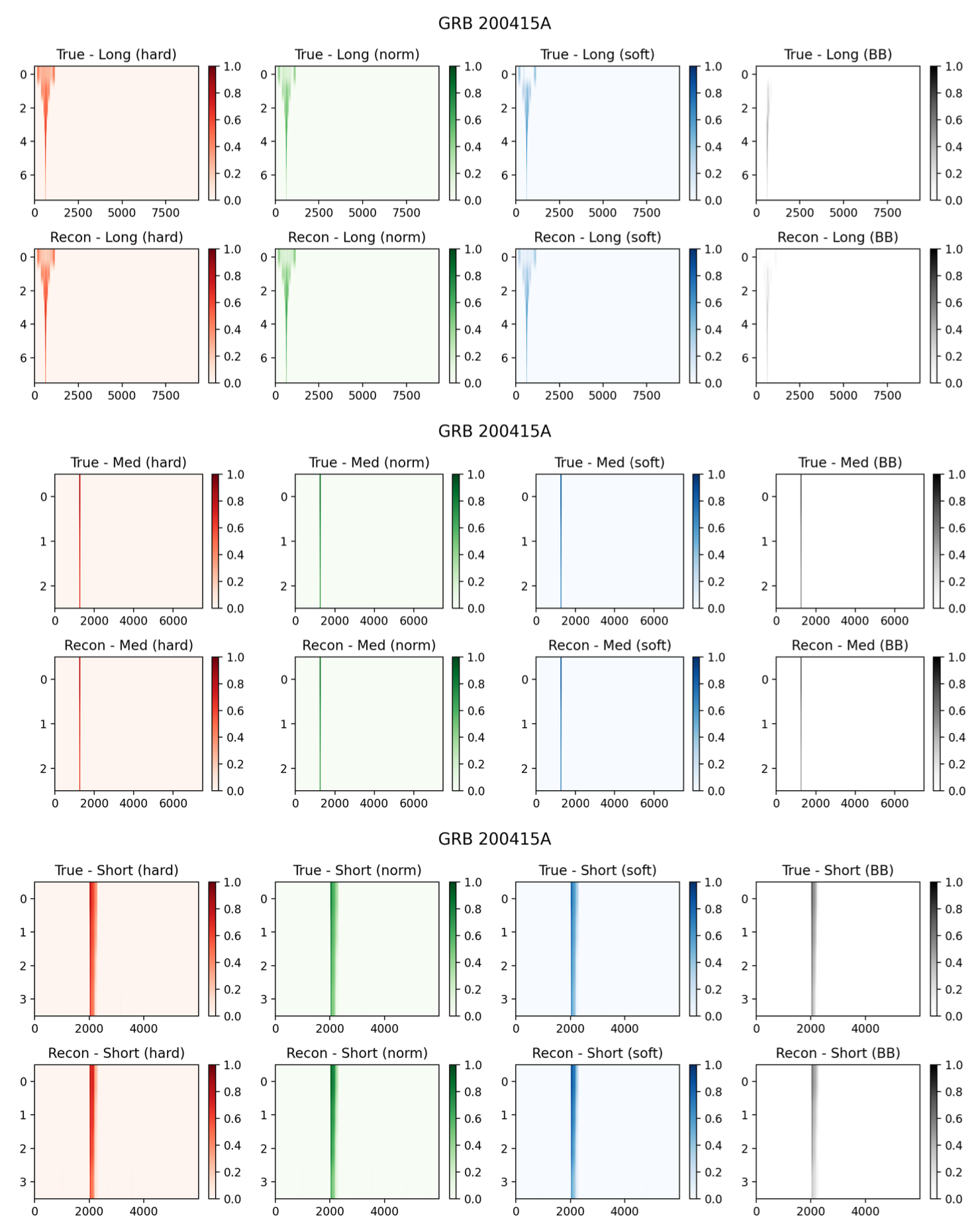}
    \caption{Reconstructed VS True waterfalls for GRB~200415A (MinVal = 10)}
    \label{fig:grb5b}
\end{figure*}

\begin{figure*}
    \centering
    \includegraphics[width=\textwidth]{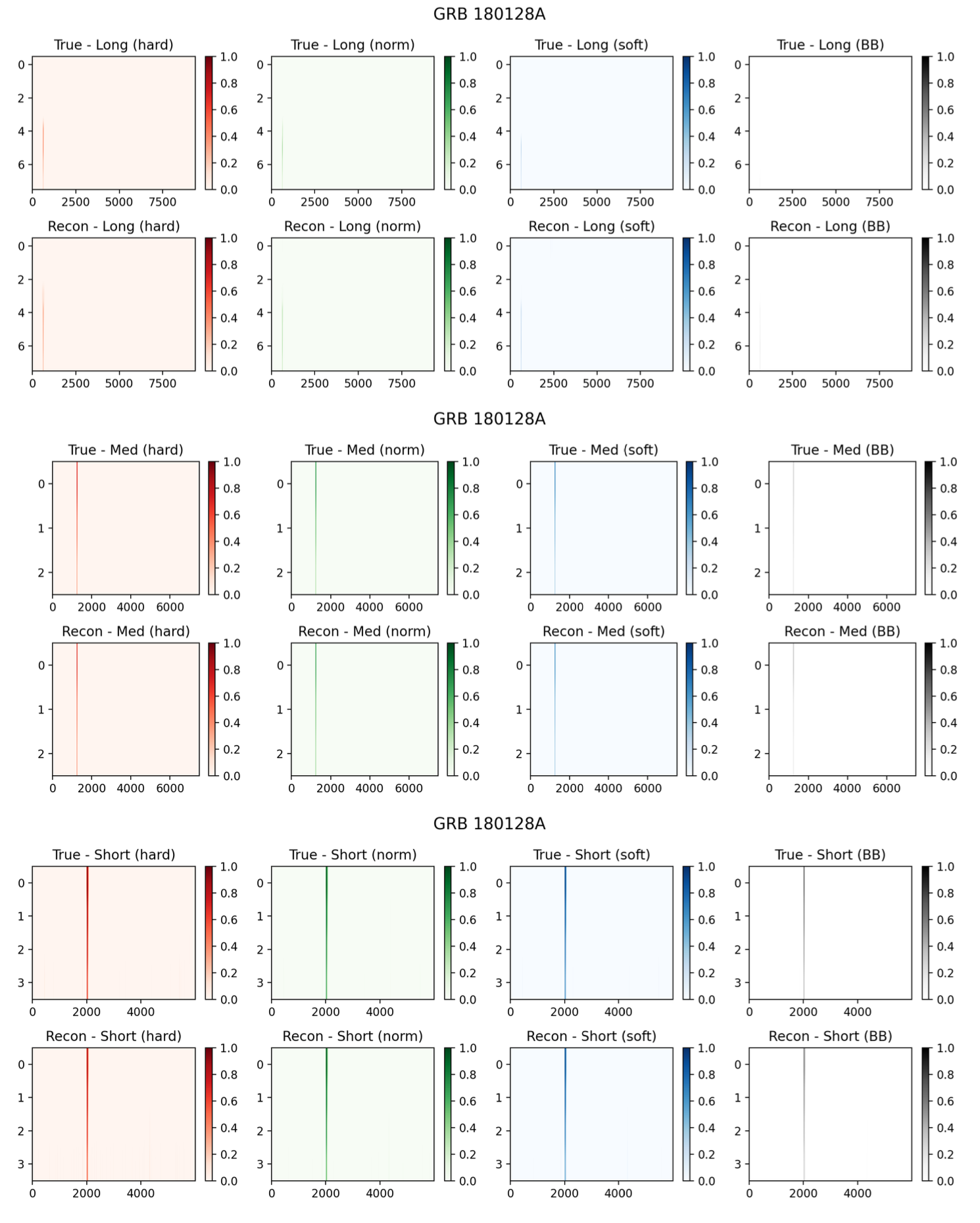}
    \caption{Reconstructed VS True waterfalls for GRB~180128A (MinVal = 10)}
    \label{fig:grb6b}
\end{figure*}

\begin{figure*}
    \centering
    \includegraphics[width=\textwidth]{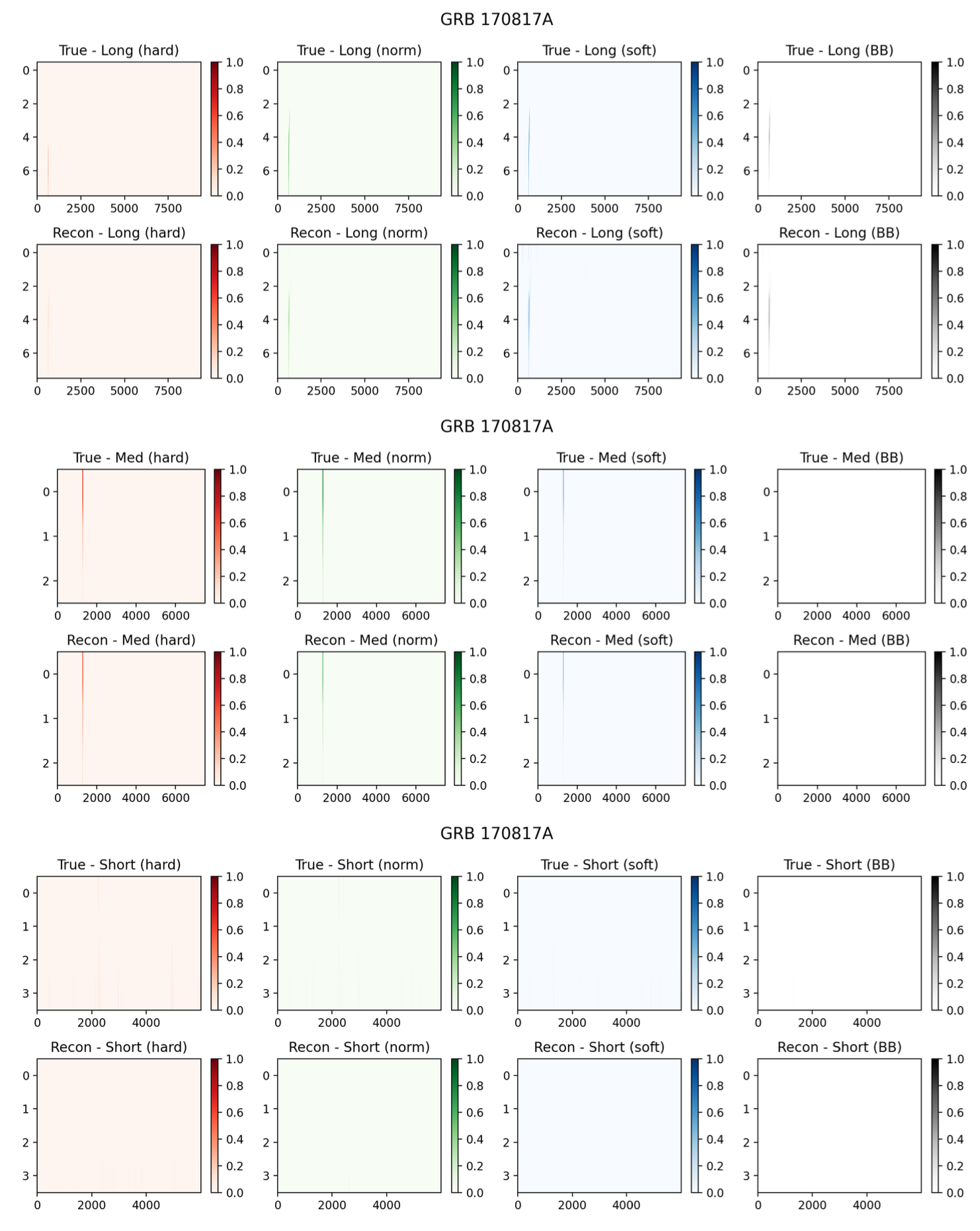}
    \caption{Reconstructed VS True waterfalls for GRB~170817A (MinVal = 10)}
    \label{fig:grb7b}
\end{figure*}



\end{document}